\documentclass[12pt]{article}
\usepackage{amssymb,amsmath,epsfig}
\usepackage{graphicx}
\allowdisplaybreaks

\begin{document}

\title{\textbf{Comparative Analysis of Holographic Dark Energy Models in
$f(R,T^2)$ Gravity}}
\author{M. Sharif$^{1,2}$\thanks{msharif.math@pu.edu.pk}~,
M. Zeeshan Gul$^{3,4,1}$\thanks{mzeeshangul.math@gmail.com,
zeeshan.gul@khazar.org}~ and I. Hashim$^1$
\thanks{imran.hashim@math.uol.edu.pk}\\
$^1$~Department of Mathematics and Statistics, The University of Lahore,\\
1-KM Defence Road Lahore-54000, Pakistan.\\
$^2$ Research Center of Astrophysics and Cosmology, Khazar
University,\\ Baku, AZ1096, 41 Mehseti Street, Azerbaijan.\\
$^3$ College of Transportation, Tongji University, Shanghai 201804,
China.\\
$^4$ Postdoctoral Station of Mechanical Engineering, Tongji
University,\\ Shanghai 201804, China.}

\date{}
\maketitle

\begin{abstract}
This study investigates the Renyi Holographic dark energy,
Sharma-Mittal Holographic dark energy and Generalized Holographic
dark energy models in the framework of $f(R,T^2)$ gravity, where $R$
denotes the Ricci scalar and $T^2$ represents the self-contraction
of the stress-energy tensor. For this purpose we employed two
horizons as infrared cut-offs, such as Hubble horizon and Ricci
horizon. The analysis is conducted for a non-interacting scenario in
a spatially flat Friedmann-Robertson-Walker universe. By considering
a specific form of this modified gravity, we reconstruct the
corresponding gravitational models based on these selected dark
energy formulations. Additionally, a stability analysis is performed
for all cases and the evolution of the equation of state parameter
is examined. Our finding indicates that the reconstructed $f(R,T^2)$
models effectively describe both the phantom and quintessence phases
of cosmic evolution, aligning with the observed accelerated
expansion of the universe. This study highlights the deep
interconnections between holographic dark energy models and modified
gravity theories, offering valuable insights into the large scale
dynamics of the cosmos.
\end{abstract}


\section{Introduction}

Einstein general relativity (GR) made a significant shift in
understanding of gravitational interactions by providing a geometric
interpretation of gravity. It explains and works accurately to study
many phenomena in the universe and has passed the solar system
tests. However, GR faces certain challenges including the
singularity inside the black hole, have caused researcher to modify
GR. Modified gravitational theories are used as an alternate
proposals to solve foundational questions while explaining dark
energy (DE) as well as the cosmic accelerated expansion and numerous
significant astronomical observations. These innovative theories add
new features in the form of curvature invariants or the geometric
components in the Einstein-Hilbert action. These modifications give
comprehensive explanation to cosmic acceleration without the need of
exotic energy components. The $f(R)$ gravity, \cite{5} which is one
of the simplest extension to GR, is a model in which the Ricci
scalar is replaced with its function in the Einstein-Hilbert action.
A comprehensive discussion of this theory and its cosmological
implications can be found in \cite{6}. Various alternative theories
and their observational constraints are examined in
\cite{z1}-\cite{z12}.

The emergence of singularities is considered as a curial problem
associated with GR, in particular at higher curvature regimes. To
overcome these problems, an innovative approach introduces
generalization of GR, known as the $f(R,T^2)$ theory or
energy-momentum squared gravity (EMSG), offering a compelling
alternative by introducing a correction term that involves the
contraction of the energy-momentum tensor
$T^2=T^{\xi\eta}T_{\xi\eta}$ \cite{7}. This novel framework explores
the dynamics of spacetime in the presence of matter and introducing
additional degrees of freedom to overcome spacetime singularity.
This correction term determines a nuanced interplay between geometry
and matter thus avoiding singularity. The incorporation of
additional non-linear terms offers explanations for enigmatic cosmic
phenomena \cite{08}. It is worthwhile to mention here that this
theory reduces to GR in vacuum with its dynamical features are
visible in high curvature regimes.

Roshan and Shojai \cite{00024} demonstrated that EMSG exhibits a
bounce during the early universe, effectively addressing primordial
singularity issues. Board and Barrow \cite{00025} employed a
specific model in this framework to explore exact solutions which
effectively demonstrate the cosmic evolution. Notably, this
innovative approach has successfully passed solar system tests
\cite{9}, enhancing its credibility. The growing interest in EMSG
among researchers arises from its significant theoretical
implications, consistency with observational data and relevance in
cosmological contexts \cite{12}-\cite{12a}. For instance, Bahamonde
et al \cite{12b} found that various models in EMSG accurately
capture the current evolutionary trends and acceleration of the
universe. Ranjit et al \cite{00028} explored solutions for matter
density and their cosmological implications. Sharif and his
collaborators \cite{z13}-\cite{z18} comprehensively examined various
aspects of this theory.

The enigmatic nature of DE and its role in the cosmic accelerated
expansion has been a central focus in cosmology. Chattopadhyay et al
\cite{029} investigated the reconstruction of the PDE model in the
framework of $f(T, T_G)$ gravity and examined that the reconstructed
model exhibits a phantom-like behavior for specific choices of model
parameters. The generalized ghost PDE has also been analyzed in
\cite{030}. Jawad and Chattopadhyay \cite{031} explored the
correspondence phenomenon in the context of modified Horava-Lifshitz
gravity and found that the cosmological parameters associated with
the reconstructed models align with present-day observational data.
Odintsov et al \cite{0034} explored specific $f(R,G)$ models to
understand their role in driving both DE-induced acceleration and
the inflationary epoch. To understand the impacts of DE on the
current acceleration of the cosmos, researchers have formulated
various DE models \cite{1}. However, majority of the existing models
has not tackled the nature of DE. This motivates the cosmologists to
pursue the novelty of a DE model grounded on the laws governing
quantum gravity and particle physics. In this respect, the
Holographic DE (HDE) model was introduced \cite{2}, offering a clear
understanding of DE while addressing theoretical challenges in the
standard cold dark matter model. In recent years, various DE models
have been established to understand the accelerated cosmic
expansion, each grounded in different theoretical foundations. Among
these, the agegraphic DE model links the energy density of DE to the
universe age, offering an intriguing approach to cosmic evolution.
Moreover, researchers have introduced novel DE models inspired by
Holographic principle and multiple entropy formalisms, leading to
the development of R\'enyi HDE (RHDE) \cite{2.001}, Tsallis HDE
\cite{2.01} and Sharma-Mittal DE (SMHDE) models \cite{2.02}.

The phenomenon of cosmic expansion has been explored through various
DE models, including RHDE and SMHDE in the framework of loop quantum
cosmology \cite{24}. Moradpour et al \cite{24.01} applied the
thermodynamic approach to HDE and RHDE to study the accelerated
expansion of the universe. Furthermore, different variants of HDE
models have been analyzed in \cite{25} with their implications in a
D-dimensional fractal universe \cite{26}. Anisotropic RHDE models in
GR has been investigated in \cite{26.01}. Sharma and Dubey \cite{27}
explored the SMHDE model in an isotropic and spatially homogeneous
flat Friedmann-Robertson-Walker (FRW) universe by considering
different values of the parameters with IR cutoff governed by the
Hubble horizon. The accretions of variants of HDE models onto
higher-dimensional Schwarzschild black hole and Morris-Thorne
wormhole have been investigated in \cite{27.01}. Shekh et al
\cite{28} examined the RHDE and SMDE models in $f(\mathcal{T},B)$
gravity (here $\mathcal{T}$ is the torsion and $B$ is the boundary
term) using the Hubble horizon as the IR cutoff. Further studies
examined the new agegraphic DE models in generalized Rastall gravity
\cite{29}. Various generalizations of the HDE paradigm have been
formulated to capture different aspects of cosmic evolution
\cite{30}. Upadhyay and Dubey \cite{30.01} diagnosed the
Sharma-Mittal HDE through state-finder and found the transition from
decelerated to accelerated expansion of the universe. The
Sharma-Mittal HDE has also been investigated in the context of
Brans-Dicke scalar-tensor gravity \cite{31}, where its cosmic
aspects has been discussed in \cite{31.01}. The study of different
DE models to study the mysterious universe has been studied in
non-metric theories has been studied in \cite{z19}-\cite{z21}.

In this study, we investigate the application of various HDE models
in the framework of $f(R,T^2)$ gravity to examine their implications
for cosmic evolution. The structure of the paper is organized as
follows. In section \textbf{2}, we derive the field equations
governing this modified gravity framework in the context of flat FRW
universe. Sections \textbf{3} and \textbf{4} focus on the analysis
of cosmic dynamics by employing the RHDE, SMHDE and GHDE models
through Hubble horizon and Ricci horizon as IR cutoff, providing a
comparative evaluation of their behavior. Finally, the key findings
of our results are given in sections \textbf{5}.

\section{The $f(R,T^2)$ theory}

The $f(R,T^2)$ gravity offers a significant development in
theoretical physics, presenting novel opportunities to examine
cosmological phenomena beyond the standard paradigm. The action of
gravity theory includes the Ricci scalar and the self-contraction of
the energy-momentum tensor. The corresponding action is formulated
as \cite{7}
\begin{equation}\label{1}
\mathcal{S}=\int d^{4}x\bigg(\frac{1}{2\kappa^{2}}
f(R,T^2)+\mathbf{L}_{m}\bigg)\sqrt{-g}.
\end{equation}
Here, the matter Lagrangian density and determinant of metric tensor
are designated by $\mathbf{L}_{m}$ and $g$ respectively, whereas,
$\kappa^{2} = 1 $ is the coupling constant. The corresponding field
equations are derived by applying the variational principle to the
action with respect to the metric tensor as
\begin{equation}\label{2}
f_{T^2}\Theta_{\xi\eta}+R_{\xi\eta}f_{R}-
\frac{1}{2}g_{\xi\eta}f+(g_{\xi\eta}\Box-
\nabla_{\xi}\nabla_{\eta})f_{R}=T_{\xi\eta},
\end{equation}
where
\begin{equation}\label{3}
\Theta_{\xi\eta}=-4\frac{\partial^{2}\mathbf{L}_{m}}{\partial
g^{\xi\eta}\partial
g^{\varphi\beta}}T^{\varphi\beta}-2\mathbf{L}_{m}
(T_{\xi\eta}-\frac{1}{2}g_{\xi\eta}T)-T T_{\xi\eta}+2
T^{\varphi}_{\xi} T_{\eta\varphi}.
\end{equation}
Here, $f_{R}=\frac{\partial f}{\partial R}$, $f_{T^2}=\frac{\partial
f}{\partial T^2},$ $\Box=\nabla^{\xi}\nabla_{\xi}$ and
$\nabla_{\xi}$ is the covariant derivative. We consider the dust
matter distribution as
\begin{equation}\label{4}
T_{\xi\eta}=\rho\mathbb{U}_{\xi}\mathbb{U}_{\eta},
\end{equation}
where $\mathbb{U}_{\mu}$ determines the four velocity and $\rho$ is
the energy density. Equation (\ref{3}) corresponding to
$\mathbf{L}_m = \rho$ turns out to be
\begin{equation}\label{5}
\Theta_{\xi\eta}=-2\rho(T_{\xi\eta}-\frac{1}{2}g_{\xi\eta}T)
-TT_{\xi\eta}+2T^{\varphi}_{\xi}T_{\varphi\eta}.
\end{equation}
Rearranging Eq.\eqref{2}, we have
\begin{eqnarray}\label{6}
G_{\xi\eta}=\frac{1}{f_{R}}(T_{\xi\eta}+T_{\xi\eta}^{DE}).
\end{eqnarray}
Here,
\begin{equation}\label{7b}
T^{DE}_{\xi\eta}=\frac{1}{2}g_{\xi\eta}(f- Rf_{R})-(g_{\xi\eta}\Box
-\nabla_{\xi}\nabla_{\eta})f_{R}- f_{T^2} \Theta_{\xi\eta}.
\end{equation}

To explore the mysterious universe, we assume the flat FRW model as
it is consistent with the cosmological principle, ensuring
homogeneity and isotropy on large scales. This model is given as
\begin{equation}\label{7c}
ds^{2}=d\mathrm{t}^{2}-{\mathbf{a}}^{2}(\mathrm{t})d\mathbf{\chi}^{2}.
\end{equation}
Here, $ d\mathbf{\chi}^{2} = d\mathbf{x}^2 + d\mathbf{y}^2 +
d\mathbf{z}^2 $ and $\mathbf{a}(t)$ represents the scale factor.
Using Eqs.(\ref{2})-(\ref{7c}), the corresponding field equations
are obtained as follows
\begin{eqnarray}\label{7d}
3H^{2}&=&\rho_{total},
\\\label{7e}
3H^{2}+2\dot{H}&=&-p_{total}.
\end{eqnarray}
The Hubble parameter $H = \frac{\dot{\mathbf{a}}}{\mathbf{a}}$
(where dot designates the derivative with respect to time) is a
significant quantity in modern cosmology that determines the cosmic
expansion rate. It serves as a curial observational tool for
understanding cosmic expansion history, the nature of DE and
deviations from standard gravity, $\rho_{total}=\rho+\rho_{DE}$,
$p_{total}=p_{DE}$ and
\begin{eqnarray}\label{8}
\rho_{DE}&=&\frac{1}{f_{R}}\bigg[\big((1- f_{R})+\rho
f_{T^2}\big)\rho+ \frac{1}{2}(f-Rf_{R}) -3H\dot{R}f_{RR}\bigg],
\\\label{9}
p_{DE}&=&\frac{1}{f_{R}}\bigg[ \frac{1}{2}(Rf_{R}-f)
+\ddot{R}f_{RR}+ \dot{R}^2f_{RRR}+ 2H\dot{R}f_{RR}\bigg].
\end{eqnarray}
The field equations are complex due to the multivariate functions
and their partial derivatives. To simplify the analysis and obtain
explicit solutions, we adopt a specific functional form for the
theory as
\begin{equation}\label{10}
f(R,T^2) = \alpha R^n +\beta T^2,
\end{equation}
where $\alpha$ and $\beta$ are non-zero constants.

This functional form is motivated by both mathematical feasibility
and physical relevance. Power-law extensions of the Ricci scalar
have been extensively explored in modified frameworks, as they are
natural generalization of GR and provide rich cosmological dynamics.
The constant $\alpha$ emerges as a coupling parameter that
determines the deviation from GR, whereas, $\beta$ plays a
significant role in exploring the influence of the $T^2$, by
introducing $\beta$, we can examine a spectrum of scenarios. Thus,
$\alpha$ and $\beta$ not only shape the overall dynamics of the
model but also provide a systematic way to investigate the interplay
between geometry and matter. Their combined effects enable the study
of a broad class of solutions that may address outstanding problems
in cosmology, such as the nature of late-time acceleration and the
role of matter-induced modifications in the evolution of the
universe. Using Eqs.(\ref{5}) and (\ref{10}) into the field
equations (\ref{8}) and (\ref{9}), we get
\begin{eqnarray}\nonumber
\rho_{DE} &=&\frac{1}{\alpha nR^{n-1}}\biggl[-3 \alpha n(n-1) H
R^{n-2} \dot{R}+\rho\big(\beta\rho +\big(1-\alpha n
R^{n-1}\big)\big)\\\label{11}&+&\frac{1}{2} \big(\beta\rho ^2-\alpha
nR^n+\alpha R^n\big)\biggr],\\\nonumber p_{DE}&=&\frac{1}{\alpha
nR^{n-1}}\biggl[\alpha n(n-1)R^{n-2} \ddot{R}+\frac{1}{2}
\biggl(-\beta\rho^2-\alpha nR^n+\alpha R^n\\\label{12}&+&2\alpha
n(n-1)H \dot{R}+\alpha
n(n-1)(n-2)R^{n-3}+\dot{R}^2\big)\biggr)\biggr].
\end{eqnarray}
In this innovative framework, the non-conserved stress-energy tensor
indicates the presence of geodesic motion of particles, leading to
additional physical effects not incorporated in standard models. The
resulting non-conserved stress-energy tensor is given by
\begin{equation}\label{11.1}
\nabla^{\xi} T_{\xi\eta}= \nabla^{\xi}\Theta_{\xi\eta}f_{T^2}
-\frac{1}{2}g_{\xi\eta}\nabla^{\xi}f(T^2).
\end{equation}
Solving this equation, we have
\begin{eqnarray}\label{12.1}
\dot{\rho}+H\rho+\dot{\rho_{DE}}+3 H\rho_{DE} (1+\omega_{DE})=
\rho^{2}\dot{f}_{T^{2}}-3H\rho^{2}f_{T^{2}}-3\dot{\rho}\rho
f_{T^{2}}.
\end{eqnarray}

In the next sections, we present a reformulation of the RHDE, SMDE
and GHDE models through Hubble horizon and Ricci horizon as IR
cut-offs. This involves revisiting their theoretical foundations,
modifying their formulations in the $f(R,T^2)$ theory and analyzing
their implications for cosmic evolution.

\section{The Hubble Horizon}

This horizon plays a significant role in gravitation and cosmology
as a natural length scale that characterizes the causal structure
and cosmic evolution. The expression for the Hubble horizon in terms
of Hubble parameter is $L_{H}=\frac{1}{H}$, it measures the distance
over which the universe expansion becomes curial in a Hubble time.
Physically, it defines the growing boundary beyond which the
particles move faster than the speed of light due to the cosmic
expansion. From theocratical point of view, it appears naturally in
the Friedmann equations as scale relating the cosmic expansion rate
to its energy density. This relationship exhibits $L_{H}$ as measure
of dynamical responses of spacetime to the total energy content.
Specifically, in HDE models, the Hubble horizon as an IR cut-off
connects large-scale cosmological dynamics with quantum field
theoretic and thermodynamic principles. Moreover, the Hubble horizon
is measured locally through the instantaneous expansion rate. This
local character makes it particularly suitable for constructing
cosmological models that aim to describe the current cosmos without
requiring knowledge of its complete history or fate.

Now, we provide a detail analysis of several well-established HDE
models in the background of Hubble horizon as IR cutoff. Using these
models, we derive the analytical expressions for the unknown
quantities in the field equations. These HDE models offer a
compelling framework for explaining the late time acceleration of
the universe. By examining these models, we aim to gain deep
insights into the nature of DE and its implications for the
evolution of the cosmic expansion.

\subsection{R\'enyi Holographic Dark Energy Model}

The R\'enyi HDE model appears from the interaction between quantum
gravity and thermodynamics, providing a modified entropy. Derived
from R\'enyi entropy, this innovative framework provides a more
generalized view on the thermodynamic properties of the universe,
potentially overcomes the coincidence problem. The Tsallis entropy
(TE) is given as
\begin{equation}\label{13}
\mathcal{S}_{\mathbf{T}}=\frac{1}{\zeta}
\Sigma_{j=1}^{X}(\mathcal{P}_{j}^{1-\zeta}-\mathcal{P}_{j}),
\end{equation}
where $\mathcal{P}_{j}$ determines the probability distribution
which satisfies $\Sigma_{j=1}^{X}\mathcal{P}_{j}=1$ and the R\'enyi
entropy is defined as
\begin{equation}\label{14}
\mathcal{S}_{RE}=\frac{1}{\zeta}\ln
\Sigma_{j=1}^{X}\mathcal{P}_{j}^{1-\zeta}.
\end{equation}
Here, $\zeta=1-\mathcal{N}$($\mathcal{N}$ is the free parameter).
Boltzmann-Gibbs entropy is recovered by substituting $\mathcal{N}=1$
in the above two equations. The R\'enyi entropy in terms of black
hole entropy is defined as
\begin{equation}\label{15}
\mathcal{S}_{RE}=\frac{1}{\zeta}\ln(1+\zeta \mathcal{S}_{BH}),
\end{equation}
Using the thermodynamic relation $\rho d\mathcal{V} \propto T
d\mathcal{S}$, where $T = \frac{1}{2\pi\mathcal{L}}$ represents the
Cai-Kim temperature associated with a system characterized by the IR
cutoff $\mathcal{L}$ and $\mathcal{S}$ denotes the horizon entropy,
we derive the energy density for the RHDE model as
\begin{equation}\label{16.0}
\rho_{RDE}=\frac{3\mathcal{C}^2}{8\mathcal{L}^2\pi(1+\zeta
\mathcal{L}^2)}.
\end{equation}
Here, $\mathcal{C}$ signifies a dimensionless quantity. For the
Hubble horizon this equation turns out to be
\begin{equation}\label{16}
\rho_{RDE}=\frac{3\mathcal{C}^2H^2}{8\mathcal{L}^2\pi(1+\frac{\zeta}{H^2})}.
\end{equation}
By specifically adopting the Hubble radius as the IR cutoff, this
formulation gives a direct connection between the thermodynamic
properties of the horizon and the evolution of DE.

We assume that the power-law scale factor naturally arises as a
solution to the Friedmann equations under specific assumptions about
the dominant energy component and is given as
\begin{equation}\label{17}
\mathbf{a}(t)=\mathbf{a}_{\circ}
\biggl(\frac{t}{t_{\circ}}\biggr)^{\delta}
e^{\eta\bigl(\frac{t}{t_{\circ}}-1\bigr)}.
\end{equation}
Here, $\delta$ and $\eta$ are non-zero positive constants and
$t_{\circ}$ determines the cosmic age, while the relationship
between time and redshift follows the Lambert function distribution
as $t(z)=\frac{s t_{\circ}}{l}g(z)$, where $s$ and $l$ are positive
constants. The parameters appearing in Eq.(\ref{17}) are carefully
selected to capture the essential features of cosmic evolution in a
viable observational framework. The constant $\mathbf{a}_{\circ}$
serves as a normalization factor, ensuring that the scale factor is
dimensionally consistent and provides a convenient initial reference
for the expansion history. The parameter $\delta$ governs the
power-law contribution to the scale factor, thereby controlling the
early-time expansion dynamics and influencing the rate of transition
toward accelerated phases. The exponential contribution determined
by $\eta$ plays a central role in generating late-time acceleration,
mimicking the effect of DE-like components. Furthermore, the
parameter $t_{\circ}$ denotes a characteristic time scale against
which cosmic evolution is measured, allowing the model to maintain
consistency with cosmological timescales inferred from observations.
The adopted parameter values are chosen to ensure a positive and
monotonically increasing scale factor and Hubble parameter,
consistent with the scenario of cosmic accelerated expansion. This
expression can be rewritten as
\begin{equation}\label{16.1}
g(z)=\text{Lambert}W\biggl(\frac{l}{s}e^{\frac{l-\ln(1+z)}{s}}\biggr).
\end{equation}
The Hubble parameter turns out to be
\begin{equation}\label{18}
H=\frac{\delta}{t}+\frac{\eta}{t_{\circ}}.
\end{equation}

Cosmic chronometers (CC) provide a significant tool to determine the
Hubble expansion rate by analyzing the relative ages of galaxies.
This approach relies on passively evolving galaxies, which are
galaxies that have stopped forming new stars and are mainly composed
of old stellar populations. By analyzing the ages and metallicities
of such galaxies at different redshifts, one can determine how fast
the cosmos is expanding. The basic idea is that the Hubble parameter
$H_{CC}(z)=\frac{1}{1+z_{CC}}\frac{dz_{CC}}{dt}$, can be estimated
from the rate of change of redshift with respect to cosmic time. In
other words, by comparing the ages of galaxies at slightly different
redshifts, we can infer how the redshift changes over time, and thus
obtain the expansion rate at that redshift \cite{31.0001}. In this
study, we use cosmic chronometer data compiled from several
observational sources, covering the redshift range $0.07\leq
z\leq1.97$ \cite{31.0002}. This dataset provides valuable,
model-independent information for constraining cosmological models
and studying the universe expansion history. Figure \textbf{1} shows
the evolution of Hubble parameter against the redshift variable. The
blue colour  represents the CC dataset and its error bars. The
figure determines that the our model is aligned with $\Lambda$CDM
model at lower redshifts. This alinement refers that the choice of
parameters significantly explains the current cosmic expansion.
\begin{figure}
\begin{center}
\epsfig{file=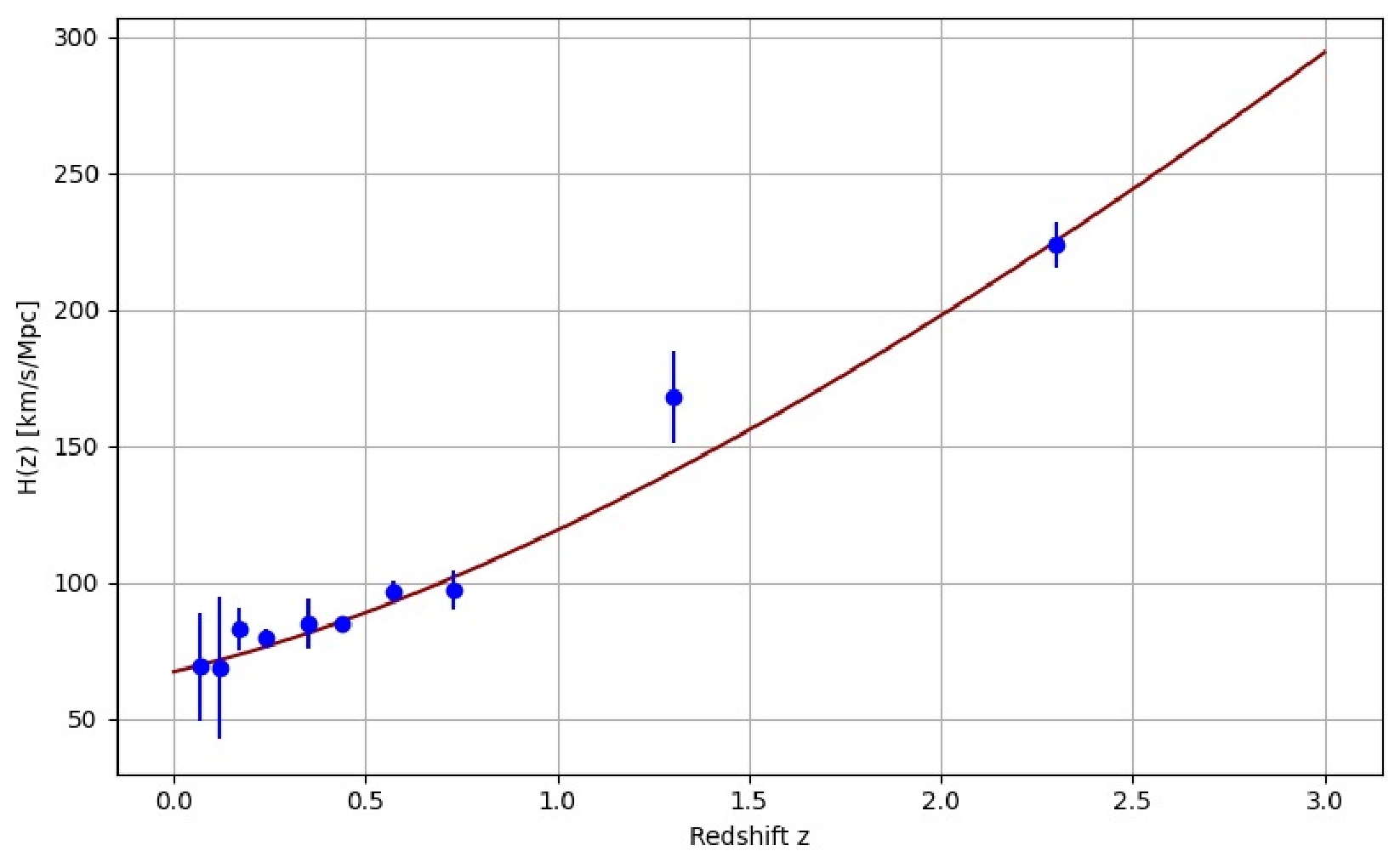,width=.5\linewidth}
\end{center}
\caption{ Trajectory of the Hubble Parameter against $z$ using CC
dataset.}
\end{figure}

The deceleration parameter $(q)$ quantifies the rate of change in
the universe expansion and is defined as
\begin{equation}\label{18.1}
q=-1-\dot{H}H^{-2}.
\end{equation}
Its value distinguishes between accelerating $(q<0)$ and
decelerating $(q>0)$ cosmic expansion, making it crucial for
understanding the impact of different energy components. From
Eqs.(\ref{18}) and (\ref{18.1}), we obtain
\begin{equation}\label{18.2}
q=-1+\frac{t_{\circ}^2\delta }{(t_{\circ}\delta+\eta t)^2}.
\end{equation}
Figure \textbf{2} presents the graphical behavior of the scale
factor, Hubble parameter and deceleration parameter as functions of
the redshift variable. The analysis is conducted for the parameter
values $\mathbf{a}_{\circ} = 10$, $l = 15$, $\delta = 10$, $\eta =
1.5$ and $t_{\circ} = 20$. The results indicate that both the scale
factor and the Hubble parameter remain positive, signifying an
accelerated cosmic expansion. This acceleration is further supported
by the negative behavior of the deceleration parameter, providing
additional confirmation of the cosmic late time accelerated
dynamics.
\begin{figure}
\epsfig{file=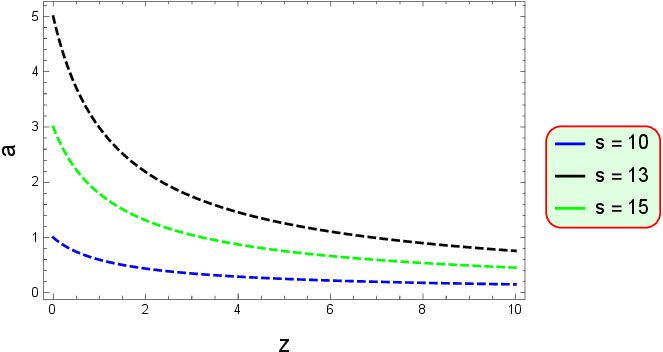,width=.5\linewidth}
\epsfig{file=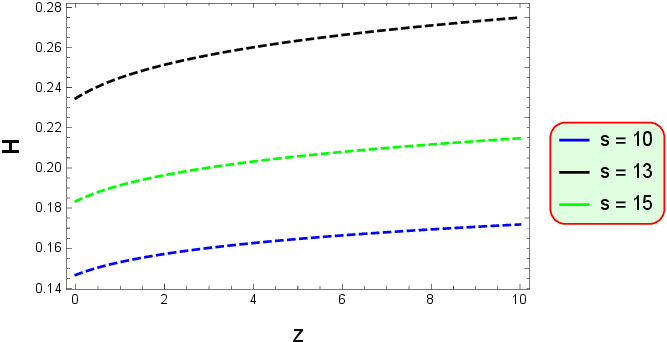,width=.5\linewidth}
\begin{center}
\epsfig{file=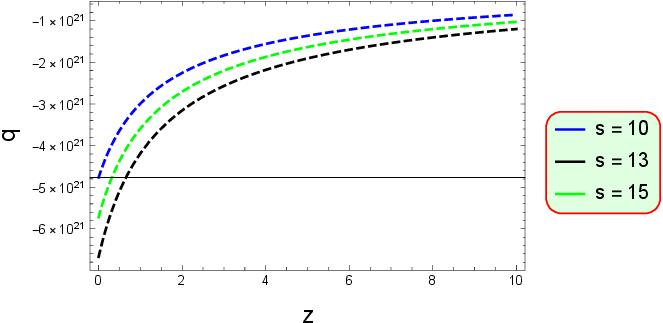,width=.5\linewidth}
\end{center}
\caption{Parametric evolution of scale factor, Hubble Parameter and
deceleration parameter.}
\end{figure}

The Ricci scalar for the flat FRW universe is obtained by using
Eq.(\ref{18}) as
\begin{equation}\label{19}
R=6\dot{H}+2H^2=\frac{6\big(t_{\circ}^2
\delta(2\delta-1)+4\delta\eta t_{\circ}
t+2\eta^2t^2\big)}{t_{\circ}^2 t^2}.
\end{equation}
Inserting Eq.(\ref{18}) into (\ref{16}), we obtain
\begin{equation}\label{20}
\rho_{RDE}=\frac{3 \mathcal{C}^2(t_{\circ}\delta+\eta t)^4}{8\pi
t_{\circ}^2 t^2 \big(t_{\circ}^2 \big(\delta^2+\pi\zeta
t^2\big)+2\delta\eta t_{\circ}t+\eta^2t^2\big)}.
\end{equation}
Using Eqs.(\ref{18}), (\ref{19}) and (\ref{20}) into (\ref{11}) and
(\ref{12}), the energy density and pressure for DE turn out to be
\begin{eqnarray}\nonumber
\rho_{DE}&=&\frac{2^{-n-6}3^{1-n}}{\alpha n}\biggl[\frac{2\eta
^2}{t_{\circ}^2}+\frac{4\delta\eta }{t_{\circ}
t}+\frac{\delta(2\delta-1)}{t^2}\biggr]^{1-n}\biggl[\alpha2^{n+6}
\biggl(\frac{6\eta^2}{t_{\circ}^2}+\frac{12\delta\eta}{t_{\circ}
t}\\\nonumber&+&\frac{3\delta(2\delta-1)}{t^2}\biggr)^n-2^{n+6}n\alpha
\biggl(\frac{6 \eta ^2}{t_{\circ}^2}+\frac{12\delta\eta}{t_{\circ}
t}+\frac{3\delta(2\delta-1)}{t^2}\biggr)^n\\\nonumber&+&\frac{9
\beta \mathcal{C}^4 (t_{\circ}\delta+\eta t)^8}{\pi^2t_{\circ}^4t^4
\biggl(\pi\alpha t_{\circ}^2t^2+(t\delta+\eta
t)^2\biggr)^2}+\biggl(2^{n+7}n\alpha(n-1)
t_{\circ}^2\delta(t_{\circ}\delta\\\nonumber&+&\eta t)
(t_{\circ}(2\delta-1)+2 \eta
t)\biggl(\frac{6\eta^2}{t_{\circ}^2}+\frac{12\delta\eta}{t_{\circ}
t}+\frac{3\delta(2\delta-1)}{t^2}\biggr)^n\biggl)
\\\nonumber&\times&\biggl(t_{\circ}^2
\delta(2\delta-1)+4t_{\circ}\delta\eta t+2\eta ^2 t^2\biggr)^{-2}
+\biggl(6\mathcal{C}^2(t_{\circ}\delta+\eta t)^4 \biggl(\pi\alpha
t_{\circ}^2\\\nonumber&-&2^{n+2}nt^2\biggl(\frac{6 \eta
^2}{b^2}+\frac{12 \delta \eta
}{t_{\circ}t}+\frac{3\delta(2\delta-1)}{t^2}\biggr)^{n-1}
\biggl(t_{\circ}^2\biggl(\delta^2+\pi\alpha
t^2\biggr)\\\nonumber&+&2t_{\circ}\delta\eta t+\eta^2
t^2\biggr)+8\pi t_{\circ}^2 t^2 \biggl(\pi\alpha
t_{\circ}^2t^2+(t_{\circ}\delta+\eta
t)^2\biggr)\\\label{21}&+&3\beta\mathcal{C}^2(t_{\circ}\delta+\eta
t)^4\biggr)\biggr)\biggl(\pi^2 t_{\circ}^4t^4 \biggl(\pi\alpha
t_{\circ}^2 t^2+(b\delta+\eta t)^2\biggr)^2\biggr)^{-1}\biggr],
\\\nonumber p_{RDE}&=&\frac{6^{-n}}{\alpha n}\biggl[\frac{2\eta^2}{t_{\circ}^2}
+\frac{4\delta\eta}{t_{\circ}t}+\frac{\delta(2\delta
-1)}{t^2}\biggr]^{1-n}\biggl[-\biggl(144\alpha\delta
n(n-1)(t_{\circ}\delta\\\nonumber&+&\eta
t)(t_{\circ}(2\delta-1)+2\eta t)\biggr)\biggl(t_{\circ}^2
t^4\biggr)^{-1}\alpha+3^{n+1}
\biggl(\frac{4\eta^2}{t_{\circ}^2}+\frac{8 \delta  \eta }{b
t}\\\nonumber&+&\frac{2\delta(2\delta-1)}{t^2}\biggr)^n\alpha
-3^{n+1}n\biggl(\frac{4\eta^2}{t_{\circ}^2}+\frac{8\delta\eta}{t_{\circ}
t}+\frac{2\delta(2\delta-1)}{t^2}\biggr)^n\\\nonumber&-&\frac{27\beta
\mathcal{C}^4(t_{\circ}\delta+\eta t)^8}{64\pi^2t_{\circ}^4t^4
\biggl(\pi\alpha t_{\circ}^2t^2+(t_{\circ}\delta+\eta
t)^2\biggr)^2}+\biggl(\alpha
t_{\circ}^4\delta^22^{n+2}n(n\\\nonumber&-&1)(n-2)
(-2t_{\circ}\delta+t_{\circ}-2\eta t)^2\biggl(\frac{6\eta
^2}{t_{\circ}^2}+\frac{12\delta\eta}{t_{\circ}t}+(3\delta(2
\delta\\\nonumber&-&1))(t^2)^{-1}\biggr)^n\biggr)\biggl(\biggl(t_{\circ}^2\delta(2\delta-1)+4t_{\circ}
\delta\eta
t+2\eta^2t^2\biggr)^3\biggr)^{-1}+\\\nonumber&&\biggl(\alpha
t_{\circ}^3\delta 2^{n+1}n(n-1)(t_{\circ}(6\delta-3)+4\eta t)
\biggl(\frac{6\eta^2}{t_{\circ}^2}+\frac{12\delta\eta}{t_{\circ}
t}+\\\label{22}&&\frac{3\delta(2\delta-1)}{t^2}\biggr)^n\biggr)\biggl(t_{\circ}^2
\delta(2\delta-1)+4t_{\circ}\delta\eta t+2\eta^2
t^2\biggr)^{-2}\biggr].
\end{eqnarray}

The equation of state (EoS) parameter, defined as
$\omega=\frac{p}{\rho}$, determines the relationship between the
matter variables in a given cosmological framework. In the
background of EMSG, this parameter is significant for exploring the
dynamic history of the cosmos. The condition where $\omega<-1$
characterizes the phantom era, in this epoch the energy density
increases with the cosmic expansion. The quintessence phase is
determined when as $\omega \in(-1,-\frac{1}{3})$, indicating a type
of DE that facilitates accelerated growth. The interplay between
these epochs, is governed by the intrinsic dynamics of the
$f(R,T^2)$ model. In this framework, the EoS parameter associated
with RHDE is determined through the expression
$\omega_{RDE}=\frac{p_{RDE}}{\rho_{RDE}}$ as
\begin{eqnarray}\nonumber
\omega_{RDE}&=&\frac{64}{3}\biggl[-\frac{144\alpha\delta
n(n-1)(t_{\circ} \delta+\eta t)(t_{\circ}(2\delta-1)+2\eta
t)}{t_{\circ}^2
t^4}\\\nonumber&+&\alpha3^{n+1}\bigl(\frac{4\eta^2}{t_{\circ}^2}+\frac{8
\delta\eta}{t_{\circ}t}+\frac{2\delta(2\delta
-1)}{t^2}\bigr)^n-3^{n+1}\alpha
n\biggl(\frac{4\eta^2}{t_{\circ}^2}\\\nonumber&+&\frac{8\delta\eta}{t_{\circ}
t}+\frac{2\delta(2\delta-1)}{t^2}\biggr)^n-\frac{27\beta
\mathcal{C}^4 (t_{\circ}\delta+\eta t)^8}{64\pi^2 t_{\circ}^4t^4
\biggl(\pi\alpha t_{\circ}^2t^2+(t_{\circ}\delta+\eta
t)^2\bigg)^2}\\\nonumber&+&\biggl(\alpha t_{\circ}^4\delta^2 2^{n+2}
n(n-1)(n-2)(-2t_{\circ}\delta+t_{\circ}-2\eta t)^2 \biggl(\frac{6
\eta^2}{t_{\circ}^2}\\\nonumber&+&\frac{12\delta\eta}{t_{\circ}t}+\frac{3\delta
(2\delta-1)}{t^2}\biggr)^n\biggr)\biggl(\biggl(t_{\circ}^2\delta(2
\delta-1)+4t_{\circ}\delta\eta t+2\eta^2
\\\nonumber&\times&t^2\biggr)^3\biggr)^{-1}+\biggl(\alpha t_{\circ}^3\delta2^{n+1}
n(n-1)(t_{\circ}(6\delta-3)+4\eta t) \biggl(\frac{6\eta
^2}{t_{\circ}^2}\\\nonumber&+&\frac{12\delta\eta}{t_{\circ}t}+\frac{3\delta(2
\delta-1)}{t^2}\biggr)^n\biggr)\biggl(t_{\circ}^2\delta(2\delta-1)+4
t_{\circ}\delta\eta
t+2\eta^2t^2\biggr)^{-2}\biggr]
\\\nonumber&&\biggl[\alpha 2^{n+6}(1-n)\biggl(\frac{6\eta^2}{t_{\circ}^2}
+\frac{12\delta\eta}{t_{\circ}t}+\frac{3\delta(2\delta
-1)}{t^2}\biggr)^n+\biggl(9\beta \mathcal{C}^4
\\\nonumber&\times&(t_{\circ}\delta+\eta
t)^8\biggr)\biggl(\pi^2t_{\circ}^4t^4\biggl(\pi\alpha
t_{\circ}^2t^2+(t_{\circ} \delta+\eta
t)^2\biggr)^2\biggr)^{-1}+\biggl(\alpha
t_{\circ}^2\\\nonumber&\times&\delta2^{n+7}
n(n-1)(t_{\circ}\delta+\eta t) (t_{\circ}(2\delta-1)+2\eta t)
\biggl(\frac{6\eta^2}{t_{\circ}^2}+\frac{12\delta\eta}{t_{\circ}
t}\\\nonumber&+&\frac{3\delta(2\delta-1)}{t^2}\biggr)^n\biggr)\biggl(\biggl(t_{\circ}^2
\delta(2\delta-1)+4t_{\circ}\delta\eta t+2\eta^2
t^2\biggr)^2\biggr)^{-1}\\\nonumber&+&\biggl(6\mathcal{C}^2(t_{\circ}
\delta+\eta t)^4\biggl(\pi\alpha t_{\circ}^2 -2^{n+2}nt^2
\biggl(\frac{6\eta^2}{t_{\circ}^2}+\frac{12\delta\eta}{t_{\circ}t}\\\nonumber&+&\frac{3
\delta(2\delta-1)}{t^2}\biggr)^{n-1}\biggl(t_{\circ}^2\biggl(\delta
^2+\pi\alpha t^2\biggr)+2t_{\circ}\delta\eta
t+\eta^2t^2\biggr)\\\nonumber&+&8 \pi t_{\circ}^2t^2\biggl(\pi\alpha
t_{\circ}^2t^2+(t_{\circ}\delta +\eta t)^2\biggr)+3\beta
\mathcal{C}^2(t_{\circ}\delta+\eta
t)^4\biggr)\biggr)\\\label{23}&\times&\biggl(\pi^2
t_{\circ}^4t^4\biggl(\pi\alpha t_{\circ}^2t^2+(t_{\circ}\delta+\eta
t)^2\biggr)^2\biggr)^{-1}\biggr]^{-1}.
\end{eqnarray}
\begin{figure}
\epsfig{file=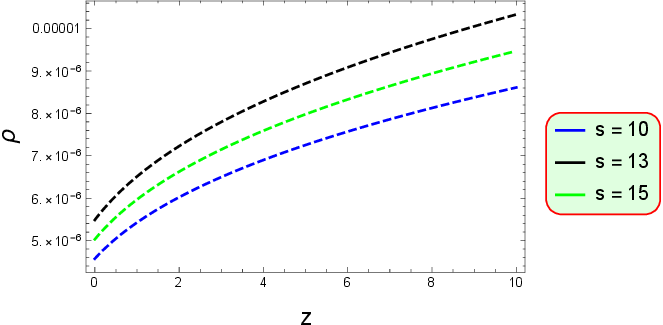,width=.5\linewidth}
\epsfig{file=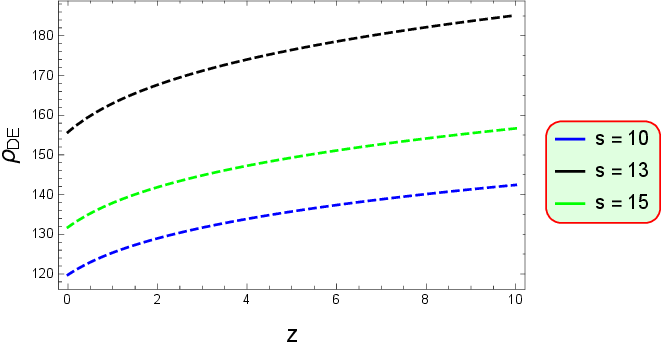,width=.5\linewidth}
\epsfig{file=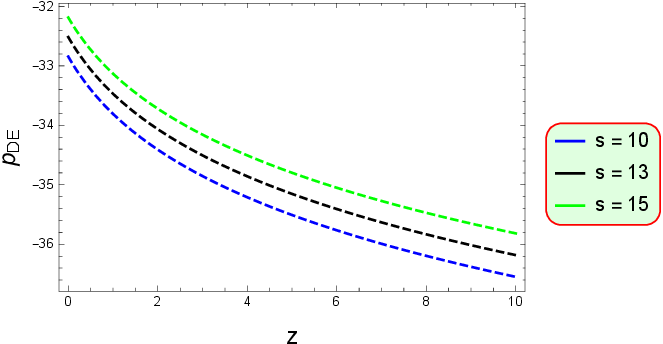,width=.5\linewidth}
\epsfig{file=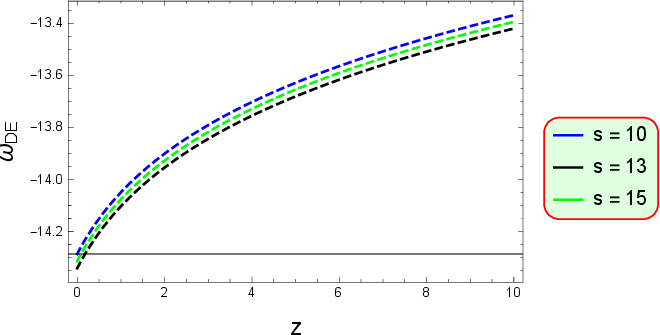,width=.5\linewidth}\caption{Graphical analysis of
$\rho$, $\rho_{DE}$, $p_{DE}$ and $\omega_{DE}$ for different
parametric values.}
\end{figure}

Figure \textbf{3} illustrates the behavior of the energy density for
RHDE, which remains positive and increases over time as the cosmos
evolved into the future. The behavior of the matter variables
associated with DE further supports the scenario of accelerated
cosmic expansion. Moreover, the EoS parameter exhibits the phantom
regime, indicating an accelerated cosmic expansion.

\subsection*{Stability Analysis}

The stability analysis of a cosmological model is vital for
determining its physical viability and consistency with
observational data. In the context of the RHDE model, evaluating
stability provides insights into the long-term behavior of cosmic
evolution. To examine the stability of the model against
perturbations, we employ the squared sound speed parameter, which is
a key diagnostic tool for determining whether perturbations in the
DE component grows or dissipates over time. A positive squared sound
speed indicates stability, while a negative value suggests an
instability in the fluctuation regime. This parameter is given as
\begin{equation}\label{23.0}
\nu_{s}^2=\frac{\dot{P}_{DE}}{\dot{\rho}_{DE}}.
\end{equation}
By substituting Eqs.(\ref{21}) and (\ref{22}) into the above
equation, we obtain the squared sound speed parameter, whose
behavior is illustrated in Figure \textbf{3}, demonstrating that it
remains positive. This indicates that the RHDE model exhibits
stability against small perturbations, ensuring that fluctuations in
the DE component do not grow over time. Consequently, the positive
squared sound speed supports the physical viability of the model in
describing the accelerated expansion of the universe.
\begin{figure}
\begin{center}
\epsfig{file=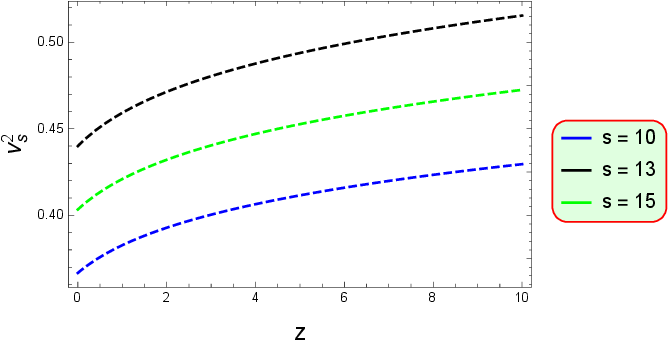,width=.5\linewidth}
\end{center}
\caption{Trajectories of squared sound speed parameter for
considered parametric values.}
\end{figure}

\subsection{Sharma-Mittal Holographic Dark Energy Model}

Sharma-Mittal entropy generalizes both Renyi and Tsallis entropies,
addressing the limitations of classical Shannon entropy in complex
systems with long range interactions and non-extensive behavior. By
incorporating two parameters, one controlling scaling from RE and
the other deformation from TE it provides great flexibility in
modeling diverse physical, statistical and cosmological systems.
This unification allows for a more comprehensive measure of
uncertainty, making it valuable in non-equilibrium dynamics, power
law distributions and multifractal structures. This entropy is given
as
\begin{equation}\label{24}
\mathcal{S}_{SMDE}=\frac{1}{1-\tau}\biggl[\biggl
(\Sigma_{j=1}^{X}\mathcal{P}_{j}^{1-\zeta}\biggr)^{\frac{1-\tau}{\zeta}}-1\biggr].
\end{equation}
Here, the free parameter $\tau=1-\mathcal{B}$. This equation takes
the form
\begin{equation}\label{25}
\mathcal{S}_{SMDE}=\frac{1}{\mathcal{B}}\biggl[\biggl(1+\frac{\chi
\mathcal{A}}{4}\biggr)^{\frac{\mathcal{B}}{\chi}}-1\biggr],
\end{equation}
where we use parametric generalized entropies
$\mathcal{S}=\frac{1}{\zeta}\ln\Sigma_{j=1}^{X}\mathcal{P}_{j}^{1-\chi}$,
$\mathcal{S}_{T}=\frac{1}{6}\Sigma_{j=1}^{X}\biggl(\mathcal{P}_{j}^{1-\chi}-\mathcal{P}_{j}\biggr)$
and the relation $\mathcal{S}_{T}=\frac{\mathcal{A}}{4}$ with
$\mathcal{A}$ represents the horizon area.

In the framework of the holographic principle, the system horizon is
intrinsically connected to both the IR and ultraviolet (UV) cutoffs.
The UV cutoff, associated with the vacuum energy density, plays a
central role in the HDE scenario. In particular, the energy density
of DE can be written as
\begin{equation}\nonumber
\rho_{DE} = \Lambda^4
\end{equation}
where $\Lambda$ denotes the UV cutoff. Cohen et al \cite{32} argued
that since the maximum energy density in the effective theory is
$\Lambda^4$, the constraint on $L$ is $L^{3}\Lambda^4\leq L
M^2_{p}$, which means that the total energy in a region of size $L$
must not exceed the mass of a black hole of the same size. This
condition enforces a connection between the UV and IR cutoffs,
ensuring that effective field theory remains valid within
holographic bounds. Although Cohen et al \cite{32} did not
explicitly propose the HDE density, their UV-IR relation was later
adapted in cosmological contexts, leading to the well known
expression for HDE in terms of the IR cutoff, which represents the
largest relevant length scale of the system. Consequently, the
expression for the holographic energy density is obtained by
enforcing these physical constraints, ensuring consistency with the
holographic principle as
\begin{equation}\label{26.0}
\rho_{SMDE}=\frac{3\mathcal{C}^2}{\mathcal{L}^4\mathcal{B}}
\biggl[\biggl(1+\chi \pi
\mathcal{L}^2\biggr)^{\frac{\mathcal{B}}{\chi}}-1\biggr].
\end{equation}
This expression for the Hubble horizon turns out to be
\begin{equation}\label{26}
\rho_{SMDE}=\frac{3\mathcal{C}^2 H^4}{\mathcal{B}}
\biggl[\biggl(1+\frac{\chi \pi}{H^2}
\biggr)^{\frac{\mathcal{B}}{\chi}}-1\biggr].
\end{equation}
Using Eqs.(\ref{18}), (\ref{19}) and (\ref{26}) into Eqs.(\ref{11})
and (\ref{12}), the expression for matter variables for DE turn out
to be
\begin{eqnarray}\nonumber
\rho_{DE}&=&\frac{2^{-n}3^{1-n}}{\alpha n}\biggl(\frac{2\eta
^2}{t_{\circ}^2}+\frac{4\delta\eta}{t_{\circ}
t}+\frac{\delta(2\delta-1)}{t^2}\biggr)^{1-n}\\\nonumber &\times&
\biggl[\alpha
6^n(1-n)\biggl(\frac{2\eta^2}{t_{\circ}^2}+\frac{4\delta
\eta}{t_{\circ}t}+\frac{\delta(2\delta-1)}{t^2}\biggr)^n
\\\nonumber&+&\frac{\beta\biggl(t_{\circ}^4\mathcal{B} t^4-3\mathcal{C}^2
(t_{\circ}\delta+\eta t)^4 \biggl(\frac{\pi t_{\circ}^2t^2\chi
}{(t_{\circ}\delta+\eta t)^2}
+1\biggr)^{B\chi}\biggr)^2}{t_{\circ}^8 \mathcal{B}^2
t^8}\\\nonumber&+&\biggl(\alpha
t_{\circ}^2\delta2^{n+1}n(n-1)(t_{\circ}\delta+\eta t)(t_{\circ}(2
\delta-1)+2\eta t)\\\nonumber&\times&
\bigg(\frac{6\eta^2}{t_{\circ}^2}+\frac{12\delta
\eta}{t_{\circ}t}+\frac{3\delta(2\delta
-1)}{t^2}\biggr)^n\biggr)\biggl(\biggl(t_{\circ}^2\delta(2\delta-1)
\\\nonumber&+&4t_{\circ}\delta
\eta t+2\eta^2t^2\biggr)^2\biggr)^{-1}-\frac{2}{t_{\circ}^8
\mathcal{B}^2 t^8}\biggl(t_{\circ}^4 \mathcal{B}t^4-3\mathcal{C}^2
(t_{\circ}\\\nonumber&\times&\delta +\eta t)^4 \biggl(\frac{\pi
t_{\circ}^2 t^2 \chi }{(t_{\circ}\delta+\eta
t)^2}+1\biggr)^{\mathcal{B}\chi}\biggr) \biggl(-t_{\circ}^4\beta
\mathcal{B}t^4+t_{\circ}^4\\\nonumber&\times&\mathcal{B}t^4+3
\beta\mathcal{C}^2 (t_{\circ}\delta+\eta t)^4\biggl(\frac{\pi
t_{\circ}^2t^2\chi}{(t_{\circ}\delta+\eta
t)^2}+1\biggr)^{\mathcal{B}\chi}\\\label{27}&+&\alpha
t_{\circ}^4\mathcal{B}
-6^{n-1}nt^4\biggl(\frac{2\eta^2}{t_{\circ}^2}+\frac{4\delta\eta}{t_{\circ}
t}+\frac{\delta(2\delta-1)}{t^2}\biggr)^{n-1}\biggr)\biggr],\\\nonumber
p_{DE}&=&\frac{6^{-n}}{\alpha n}\biggl(\frac{2\eta
^2}{t_{\circ}^2}+\frac{4\delta\eta}{t_{\circ}t}+\frac{\delta(2
\delta-1)}{t^2}\biggr)^{1-n}\\\nonumber&\times& \biggl[-\frac{144
\alpha\delta n(n-1)(t_{\circ}\delta+\eta t) (t_{\circ}(2\delta-1)+2
\eta t)}{t_{\circ}^2 t^4}\\\nonumber&+&\biggl(\alpha t_{\circ}^4
\delta^22^{n+2}n(n-1)(n-2)(t_{\circ}(2\delta-1)+2\eta
t)^2\\\nonumber&\times&
\biggl(\frac{6\eta^2}{t_{\circ}^2}+\frac{12\delta\eta}{t_{\circ}t}+\frac{3
\delta(2\delta-1)}{t^2}\biggr)^n\biggr)\biggl(\biggl(t_{\circ}^2
\delta(2\delta-1)\\\nonumber&+&4t_{\circ}\delta\eta t+2\eta^2
t^2\biggr)^3\biggr)^{-1}+\biggl(\alpha t_{\circ}^3 \delta2^{n+1}
n(n-1)(t_{\circ}\\\nonumber&\times&(6\delta-3)+4\eta t)
\biggl(\frac{6\eta
^2}{t_{\circ}^2}+\frac{12\delta\eta}{t_{\circ}t}+\frac{3\delta(2
\delta-1)}{t^2}\biggr)^n\biggr)\\\nonumber&\times&\biggl(\biggl(t_{\circ}^2\delta(2\delta-1)+4t_{\circ}
\delta\eta
t+2\eta^2t^2\biggr)^2\biggr)^{-1}+\biggl(3\\\nonumber&\times&\biggl(\alpha
t_{\circ}^8\mathcal{B}^26^nt^8\biggl(\frac{2\eta^2}{t_{\circ}^2}+\frac{4
\delta\eta}{t_{\circ}t}+\frac{\delta(2\delta
-1)}{t^2}\biggr)^n-\alpha
t_{\circ}^8\\\nonumber&\times&\mathcal{B}^26^nnt^8
\biggl(\frac{2\eta^2}{t_{\circ}^2}+\frac{4\delta\eta }{t_{\circ}
t}+\frac{\delta(2\delta
-1)}{t^2}\biggr)^n\\\nonumber&-&\beta\biggl(t_{\circ}^4
\mathcal{B}t^4-3\mathcal{C}^2 (t_{\circ}\delta+\eta t)^4
\biggl(\frac{\pi t_{\circ}^2t^2\chi }{(t_{\circ}\delta+\eta
t)^2}\\\label{28}&+&1\biggr)^{\mathcal{B}\chi}\biggr)^2\biggr)\biggr)\biggl(t_{\circ}^8
\mathcal{B}^2t^8\biggr)^{-1}\biggr].
\end{eqnarray}
The corresponding EoS parameter for the SMHDE is given as
\begin{eqnarray}\nonumber
\omega_{DE}&=&\biggl[-\frac{144\alpha\delta n(n-1)(t_{\circ}\delta
+\eta t) (t_{\circ}(2\delta-1)+2\eta t)}{t_{\circ}^2
t^4}\\\nonumber&+&\biggl(\alpha t_{\circ}^4
\delta^22^{n+2}n(n-)(n-2)(t_{\circ}(2\delta-1)+2\eta t)^2
\biggl(\frac{6\eta
^2}{t_{\circ}^2}\\\nonumber&+&\frac{12\delta\eta}{t_{\circ}t}+\frac{3\delta(2
\delta-1)}{t^2}\biggr)^n\biggr)\biggr(\biggl(t_{\circ}^2\delta(2\delta-1)+4t_{\circ}
\delta\eta
t+\\\nonumber&&2\eta^2t^2\biggr)^3\biggr)^{-1}+\biggl(\alpha
t_{\circ}^3 \delta2^{n+1}n(n-1)(t_{\circ}(6\delta-3)+4\eta
t)\\\nonumber&\times&\biggl(\frac{6\eta
^2}{t_{\circ}^2}+\frac{12\delta\eta}{t_{\circ}t}+\frac{3\delta(2
\delta-1)}{t^2}\biggr)^n\biggr)\biggl(\biggr(t_{\circ}^2\delta(2\delta-1)+4t_{\circ}
\\\nonumber&\times&
\delta\eta t+2\eta^2t^2\biggr)^2\biggr)^{-1}+\frac{3}{t_{\circ}^8
\mathcal{B}^2t^8}\biggl(\alpha t_{\circ}^8\mathcal{B}^26^nt^8
\biggl(\frac{2\eta^2}{t_{\circ}^2}+\frac{4\delta\eta}{t_{\circ}
t}\\\nonumber&+&\frac{\delta(2\delta-1)}{t^2}\biggr)^n-\alpha
t_{\circ}^8
\mathcal{B}^26^nnt^8\biggl(\frac{2\eta^2}{t_{\circ}^2}+\frac{4
\delta\eta}{t_{\circ}t}+\frac{\delta(2\delta
-1)}{t^2}\biggr)^n\\\nonumber&-&\beta\biggl(t_{\circ}^4\mathcal{B}t^4-3\mathcal{C}^2
(t_{\circ}\delta+\eta t)^4 \biggl(\frac{\pi t_{\circ}^2t^2\chi
}{(t_{\circ}\delta+\eta t)^2}+1\biggr)^{\mathcal{B}\chi
}\biggr)^2\biggr)^{-1}\biggr]\\\nonumber&\times&\biggl[3\biggl(
6^n\alpha(1-n)\biggl(\frac{2\eta^2}{t_{\circ}^2}+\frac{4\delta\eta}{t_{\circ}
t}+\frac{\delta(2\delta-1)}{t^2}\biggr)^n+\biggl(\alpha t_{\circ}^2
\delta \\\nonumber&\times&2^{n+1}n(n-1)(t_{\circ}\delta+\eta
t)(t_{\circ}(2\delta-1)+2 \eta
t)\biggr(\frac{6\eta^2}{t_{\circ}^2}+\frac{12\delta\eta}{t_{\circ}
t}\\\nonumber&+&\frac{3\delta(2\delta-1)}{t^2}\biggr)^n\biggr)\biggl(\biggl(t_{\circ}^2
\delta (2\delta-1)+4 t_{\circ}\delta\eta t+2\eta^2
t^2\biggr)^2\biggr)^{-1}\\\nonumber&+&\frac{\beta\biggl(t_{\circ}^4\mathcal{B}t^4-3\mathcal{C}^2
(t_{\circ}\delta+\eta t)^4 \biggl(\frac{\pi t_{\circ}^2t^2
\chi}{(t_{\circ}\delta+\eta t)^2}+1\biggr)^{\mathcal{B}\chi
}\biggr)^2}{t_{\circ}^8 \mathcal{B}^2
t^8}\\\nonumber&-&\frac{2}{t_{\circ}^8 \mathcal{B}^2 t^8}
\biggl(t_{\circ}^4 \mathcal{B}t^4-3\mathcal{C}^2(t_{\circ}\delta
+\eta t)^4 \biggl(\frac{\pi t_{\circ}^2t^2\chi}{(t_{\circ}\delta
+\eta
t)^2}\\\nonumber&+&1\biggr)^{\mathcal{B}\chi}\biggr)\biggl(-t_{\circ}^4
\beta\mathcal{B}t^4+t_{\circ}^4\mathcal{B}t^4+3\beta \mathcal{C}^2
(t_{\circ}\delta+\eta t)^4\\\nonumber&\times&\biggl(\frac{\pi
t_{\circ}^2t^2 \chi}{(t_{\circ}\delta+\eta
t)^2}+1\biggr)^{\mathcal{B}\chi }-\alpha t_{\circ}^4
\mathcal{B}6^{n-1}nt^4\biggl(\frac{2\eta
^2}{t_{\circ}^2}\\\label{29}&+&\frac{4\delta\eta}{t_{\circ}t}+\frac{\delta(2
\delta -1)}{t^2}\biggr)^{n-1}\biggr)\biggr)\biggr]^{-1}.
\end{eqnarray}
\begin{figure}
\epsfig{file=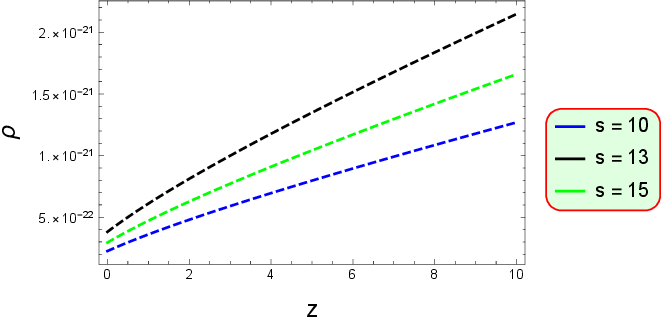,width=.5\linewidth}
\epsfig{file=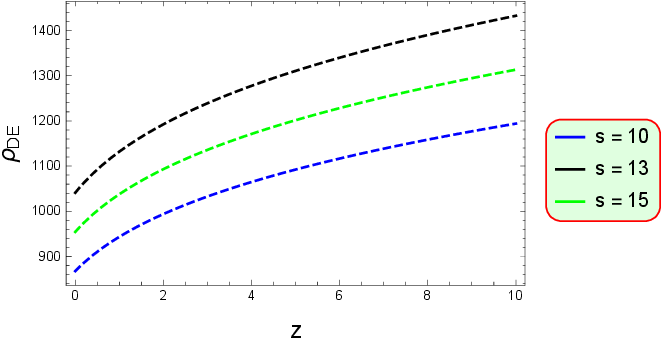,width=.5\linewidth}
\epsfig{file=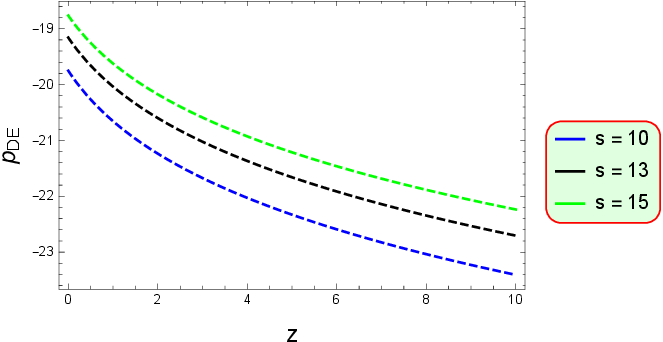,width=.5\linewidth}
\epsfig{file=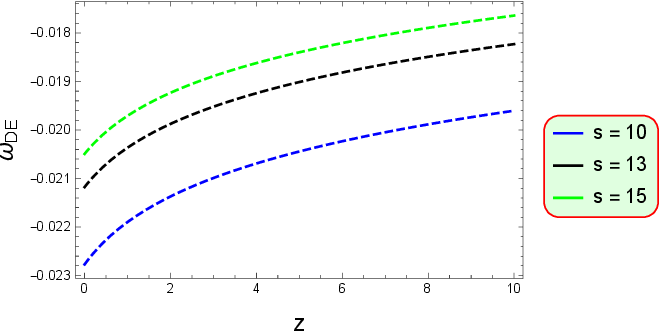,width=.5\linewidth}
\begin{center}
\epsfig{file=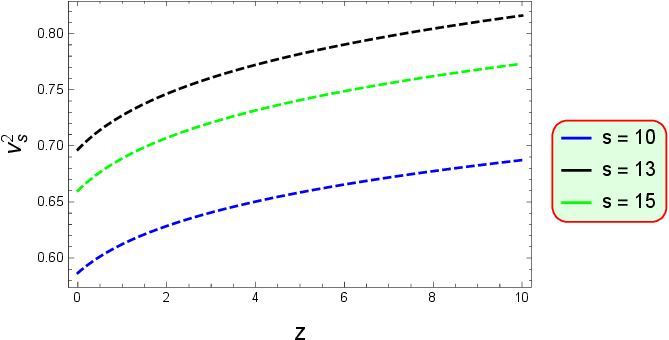,width=.5\linewidth} \end{center} \caption{Graphical
analysis of $\rho$, $\rho_{DE}$, $p_{DE}$, $\omega_{DE}$ and
$\nu_{s}^2$.}
\end{figure}

Figure \textbf{5} demonstrates that the energy density of the SMHDE
model remains positive and increases in the late-time evolution of
the universe. The behavior of energy density and pressure in this DE
model indicates an accelerating cosmic expansion, as the energy
density remains positive while the pressure takes negative values.
Additionally, the EoS parameter falls within the quintessence
regime, reinforcing the model role in driving the accelerated
expansion of the universe. Furthermore, the stability analysis,
conducted through the squared sound speed parameter, confirms that
the model remains dynamically stable throughout its evolution.

\subsection{Generalized Holographic Dark Energy Model}

The GHDE extends the standard HDE framework to address the
fine-tuning and coincidence problem associated with the cosmological
constant. By incorporating modified IR cutoffs interactions with
dark matter or higher-order curvature corrections, GHDE provides a
more flexible description of DE dynamics. This generalization allows
for better consistency with observational constraints and also
aligning with modified gravity theories. Thus, the GHDE serves as a
viable alternative framework for exploring the nature of DE. A novel
entropy function incorporating three parameters has been proposed,
capable of generalizing various existing entropy measures, including
the standard Boltzmann-Gibbs entropy, RE and TE \cite{32}. This
entropy characterized by the free parameters $\mu$, $\nu$ and
$\gamma$ is defined as
\begin{equation}
\mathcal{S}_{GHDE}[\mu,\nu,\gamma]=\frac{1}{\gamma}\biggl[\biggl(1+\frac{\mu}{\nu}
\mathcal{S}_{BH}\biggr)^{\nu}-1\biggr].
\end{equation}
The energy density for GHDE is obtained using black hole energy
density relation $\rho=24\pi \mathcal{C}^2 G \mathcal{L}^2$, with
Hubble radius $r_{H}=\frac{1}{H}$ in generalized entropy as
\begin{equation}
\rho_{GHDE}=\frac{3\mathcal{C}^2H^2}{8\pi\mathcal{B}}\bigg[\biggl(\frac{b\pi}{a
H^2}+1\biggr)^a-1\biggr],
\end{equation}
where $a$ and $b$ are non-negative parameters. The energy density
and pressure for this DE model are obtained using this equation and
Eq.(\ref{19}) into Eqs.(\ref{11}) and (\ref{12}) as
\begin{eqnarray}\nonumber
\rho_{DE}&=&\frac{2^{-n-6}3^{1-n}}{\alpha n} \biggl(\frac{2\eta
^2}{t_{\circ}^2}+\frac{4\delta\eta}{t_{\circ}t}+\frac{\delta(2
\delta-1)}{t^2}\biggr)^{1-n}
\\\nonumber&\times&\biggl[\biggl(\alpha2^{n+6}(n-1)
\biggl(\frac{6\eta^2}{t_{\circ}^2}+\frac{12\delta\eta}{t_{\circ}t}
+\frac{3\delta(2\delta-1)}{t^2}\biggr)^n\\\nonumber&\times&
\biggl(t_{\circ}^4\delta^2 (2\delta-1)(-2\delta+2 n+1)-2
t_{\circ}^3\delta\eta
t\biggl(8\delta^2\\\nonumber&-&4\delta(n+1)+n\biggr)+4
t_{\circ}^2\delta\eta^2t^2(-6\delta
+n+1)-\\\nonumber&&16t_{\circ}\delta\eta^3 t^3-4\eta^4
t^4\biggr)\biggr)\biggl(\biggl(t_{\circ}^2\delta(2\delta-1)+4t_{\circ}\delta
\eta t\\\nonumber&+&2\eta^2t^2\biggr)^2\biggr)^{-1}+\frac{27\beta
\mathcal{C}^4(t_{\circ}\delta+\eta t)^4}{\pi^2
t_{\circ}^4\mathcal{B}^2 t^4}\biggl(\biggl(\frac{\pi a t_{\circ}^2
t^2}{b(t_{\circ}\delta+\eta
t)^2}\\\nonumber&+&1\biggr)^{b}-1\biggr)^2+\frac{8\mathcal{C}^2(t_{\circ}
\delta +\eta t)^2}{\pi t_{\circ}^2
\mathcal{B}t^2\biggl(t_{\circ}^2\delta(2\delta-1)+4t_{\circ}\delta
\eta t+2\eta^2
t^2\biggr)}\\\nonumber&\times&\biggl(t_{\circ}^2\biggl(6\delta(2\delta-1)-\alpha
6^nnt^2\biggl(\frac{2\eta^2}{t_{\circ}^2}+\frac{4\delta\eta}{t_{\circ}
t}+\frac{\delta
(2\delta-1)}{t^2}\biggl)^n\biggr)\\\label{30}&+&24t_{\circ}\delta
\eta t+12\eta^2t^2 \biggl(\biggl(\frac{\pi a t_{\circ}^2 t^2}{b
(t_{\circ}\delta+\eta t)^2}+1\biggr)^{b
}-1\biggr)\biggr)\biggr],
\\\nonumber p_{DE}&=&\frac{6^{-n}}{\alpha n}\biggl(\frac{2\eta^2}{t_{\circ}^2}
+\frac{4\delta\eta}{t_{\circ}t}+\frac{\delta(2\delta
-1)}{t^2}\biggr)^{1-n}\biggl[-\frac{144\alpha}{b^2t^4}\delta
n\\\nonumber&\times&(n-1)(t_{\circ}\delta+\eta t)
(t_{\circ}(2\delta-1)+2\eta t)+\biggl(\alpha t_{\circ}^4\delta^2
2^{n+2}\\\nonumber&\times&n(n-1)(n-2)(t_{\circ}(2\delta-1)+2\eta
t)^2 \biggl(\frac{6
\eta^2}{t_{\circ}^2}+\frac{12\delta\eta}{t_{\circ}t}\\\nonumber&+&\frac{3\delta
(2\delta-1)}{t^2}\biggr)^n\biggr)\biggl(\biggl(t_{\circ}^2
\delta(2\delta-1)+4t_{\circ}\delta\eta t+2\eta^2
t^2\biggr)^3\biggr)^{-1}\\\nonumber&+&\biggl(\alpha t_{\circ}^3
\delta 2^{n+1}n(n-1)(t_{\circ}(6\delta-3)+4\eta t) \biggl(\frac{6
\eta^2}{t_{\circ}^2}+\frac{12\delta\eta}{t_{\circ}t}\\\nonumber&+&\frac{3\delta
(2\delta-1)}{t^2}\biggr)^n\biggr)\biggl(\biggl(t_{\circ}^2\delta(2\delta-1)+4t_{\circ}
\delta\eta t+2\eta^2t^2\biggr)^2\biggr)^{-1}
\\\nonumber&+&\biggl(3\biggl(-9\beta\mathcal{C}^4(t_{\circ}\delta+\eta t)^4
\biggl(\biggl(\frac{\pi a t_{\circ}^2t^2}{b(t_{\circ}\delta+\eta
t)^2}+1\biggr)^{b}-1\biggr)^2\\\nonumber&+&\pi^2 a t_{\circ}^4
\mathcal{B}^2 2^{n+6}t^4
\biggl(\frac{6\eta^2}{t_{\circ}^2}+\frac{12\delta\eta}{t_{\circ}t}
+\frac{3\delta(2\delta-1)}{t^2}\biggr)^n\\\nonumber&-&\pi^2\alpha
t_{\circ}^4 \mathcal{B}^2 2^{n+6}nt^4
\biggl(\frac{6\eta^2}{t_{\circ}^2}
+\frac{12\delta\eta}{t_{\circ}t}+\frac{3\delta(2\delta
-1)}{t^2}\biggr)^n\biggr)\biggr)\\\label{31}&\times&\biggl(64\pi^2t_{\circ}^4\mathcal{B}^2
t^4\biggr)^{-1}\biggr].
\end{eqnarray}
The EoS parameter for this DE model is expressed as
\begin{eqnarray}\nonumber
\omega_{DE}&=&\frac{64}{3}\biggl[-\frac{144\alpha\delta
n(n-1)(t_{\circ}\delta+\eta t) (t_{\circ}(2\delta-1)+2\eta
t)}{t_{\circ}^2t^4}\\\nonumber&+&\biggl(\alpha t_{\circ}^4\delta^2
2^{n+2}n(n-1)(n-2)(t_{\circ}(2\delta-1)+2\eta t)^2\biggl(\frac{6
\eta^2}{t_{\circ}^2}\\\nonumber&+&\frac{12\delta\eta}{t_{\circ}t}+\frac{3\delta
(2\delta-1)}{t^2}\biggr)^n\biggr)\biggl(\biggl(t_{\circ}^2\delta(2\delta-1)+4t_{\circ}\delta
\eta
t+2\eta^2\\\nonumber&\times&t^2\biggr)\biggr)^{-3}+\biggl(\alpha
t_{\circ}^3\delta2^{n+1}n(n-1)(t_{\circ}(6\delta-3)+4\eta t)
\biggl(\frac{6\eta^2}{t_{\circ}^2}\\\nonumber&+&\frac{12\delta\eta}{t_{\circ}
t}+\frac{3\delta(2\delta-1)}{t^2}\biggr)^n\biggr)\biggl(t_{\circ}^2
\delta(2\delta-1)+4 t_{\circ}\delta\eta t+2\eta^2
t^2\biggr)^{-2}
\\\nonumber&+&\biggl(3\biggl(-9\beta \mathcal{C}^4 (t_{\circ}\delta
+\eta t)^4\biggl(\biggl(\frac{\pi\alpha t_{\circ}^2 t^2}{b(a\delta
+\eta t)^2}+1\biggr)^{b}-1\biggr)^2+\pi^2\\\nonumber&\times&\alpha
t_{\circ}^4
\mathcal{B}^22^{n+6}t^4\biggl(\frac{6\eta^2}{t_{\circ}^2}+\frac{12
\delta\eta}{t_{\circ}t}+\frac{3\delta(2\delta -1)}{t^2}\biggr)^n-\pi
^2\alpha t_{\circ}^4
\mathcal{B}^22^{n+6}\\\nonumber&\times&nt^4\biggl(\frac{6\eta
^2}{t_{\circ}^2}+\frac{12\delta\eta}{t_{\circ}t}+\frac{3\delta(2
\delta-1)}{t^2}\biggr)^n\biggr)\biggr)\biggl(64\pi^2 t_{\circ}^4
\mathcal{B}^2 t^4\biggr)^{-1}\biggr]\\\nonumber&\times&
\biggl[3\biggl(\biggl(\alpha2^{n+6}(n-1)\biggl(\frac{6\eta
^2}{t_{\circ}^2}+\frac{12\delta\eta}{t_{\circ}t}+\frac{3\delta(2
\delta-1)}{t^2}\biggr)^n\biggl(t_{\circ}^4\delta^2\\\nonumber&\times&(2\delta-1)(-2
\delta+2n+1)-2t_{\circ}^3\delta\eta t\biggl(8\delta^2-4\delta
(n+1)+n\biggr)\\\nonumber&+&4t_{\circ}^2\delta\eta^2t^2(-6\delta+n+1)-16t_{\circ}
\delta\eta^3t^3-4\eta^4
t^4\biggr)\biggr)\biggl(b^2\delta(2\delta\\\nonumber&-&1)+4b
\delta\eta t+2\eta^2
t^2\biggr)^{-2}\biggr)+\biggl(8\mathcal{C}^2(t_{\circ}\delta+\eta
t)^2\biggl(t_{\circ}^2\biggl(6\delta(2\delta\\\nonumber&-&1)-\alpha6^nnt^2
\biggl(\frac{2\eta^2}{t_{\circ}^2}+\frac{4\delta\eta}{t_{\circ}
t}+\frac{\delta(2\delta -1)}{t^2}\biggr)^n\biggr)+24t_{\circ}\delta
\eta t\\\nonumber&+&12\eta^2t^2\biggr)\biggl(\biggl(\frac{\pi\alpha
t_{\circ}^2 t^2}{b (a\delta+\eta
t)^2}+1\biggr)^{b}-1\biggr)\biggr)\biggl(\pi t_{\circ}^2\mathcal{B}
t^2\\\nonumber&\times&\biggl(t_{\circ}^2\delta(2\delta-1)+4t_{\circ}\delta\eta
t+2\eta ^2 t^2\biggr)+\frac{27\beta \mathcal{C}^4}{\pi^2
t_{\circ}^4\mathcal{B}^2 t^4}(t_{\circ}\\\label{32}&\times&\delta
+\eta t)^4\biggl(\biggl(\frac{\pi\alpha t_{\circ}^2t^2}{b(a\delta
+\eta t)^2}+1\biggr)^{b}-1\biggr)^2\biggr)\biggr]^{-1}.
\end{eqnarray}
\begin{figure}
\epsfig{file=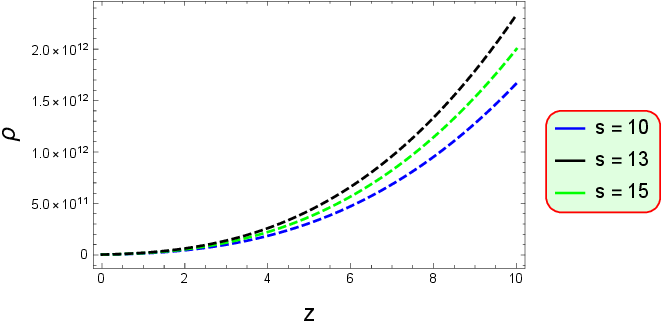,width=.5\linewidth}
\epsfig{file=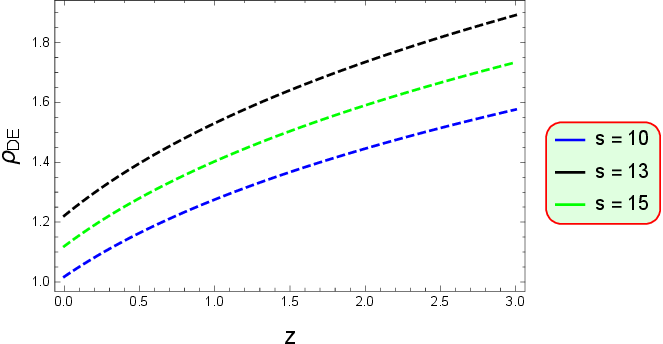,width=.5\linewidth}
\epsfig{file=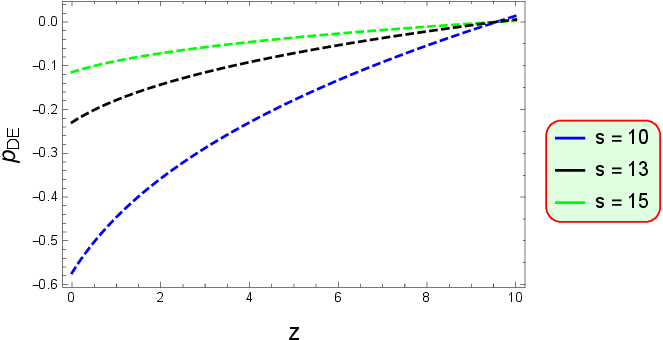,width=.5\linewidth}
\epsfig{file=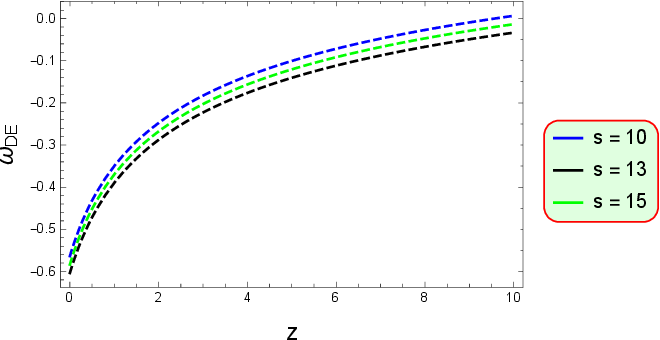,width=.5\linewidth}
\begin{center}
\epsfig{file=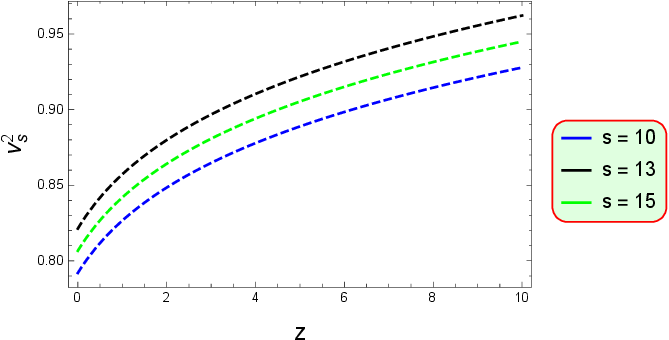,width=.5\linewidth}
\end{center}
\caption{Parametric behavior of $\rho$, $\rho_{DE}$, $p_{DE}$,
$\omega_{DE}$ and $\nu_{s}^2$.}
\end{figure}

Figure \textbf{6} determines that the energy density of the GHDE
model remains positive and increases over time, contributing to the
universe accelerated expansion. This behavior is further supported
by the matter variables for the DE associated with the model. The
EoS parameter transitions from the phantom regime in the early
universe to the quintessence regime at late times, highlighting its
evolving nature in cosmic dynamics. Additionally, stability analysis
using the squared sound speed parameter confirms the model dynamical
consistency throughout its evolution.

\section{Ricci Horizon}

The Ricci horizon $L_{R}=\sqrt{H^2+(\dot{H})^2}$ is introduced as a
dynamically motivated length scale that depends on both the Hubble
parameter $H$ and its time derivative  $\dot{H}$ , thereby
incorporating the effects of cosmic expansion and its evolution. It
is directly related to the Ricci scalar curvature of spacetime,
making it a natural geometric quantity for describing the
large-scale dynamics of the universe. In holographic DE models, the
Ricci horizon serves as a physically consistent infrared cutoff that
connects the dark energy density with spacetime curvature. Unlike
the event horizon, it is a local quantity and does not rely on the
future evolution of the scale factor, thus avoiding causality
problems while maintaining compatibility with observational
cosmology.

\subsection{R\'enyi Holographic Dark Energy Model}

The R'enyi HDE density corresponding to the Ricci horizon can be
formulated by employing Eq. (\ref{16.0}) together with the Ricci
length scale $L_{R}=\sqrt{H^2+(\dot{H})^2}$ as
\begin{equation}\label{120}
\rho_{RDE}=\frac{3 \mathcal{C}^2(t_{\circ}\delta+\eta t)^4}{8\pi
t_{\circ}^2 t^2 \bigl(t_{\circ}^2\bigl(\delta^2+\pi\alpha
t^2\bigr)+2t_{\circ}\delta \eta t+\eta^2t^2\bigr)}.
\end{equation}
The expression for the matter variables for DE are obtained using
Eqs.(\ref{18}), (\ref{19}) and (\ref{120}) into (\ref{11}) and
(\ref{12}), as
\begin{eqnarray}\nonumber
\rho_{DE}&=&\frac{2^{-n-6}3^{1-n}}{\alpha n}\bigg(\frac{2 \eta
^2}{t_{\circ}^2}+\frac{4 \delta\eta}{
t_{\circ}t}+\frac{\delta(2\delta
-1)}{t_{\circ}^2t^2}\bigg)^{1-n}\\\nonumber &\times& \bigg[\alpha
2^{n+6} \bigg(\frac{6\eta^2}{t_{\circ}^2}+\frac{12\delta\eta}{
t_{\circ}t}+\frac{3\delta(2\delta-1)}{t^2}\bigg)^n\\\nonumber &-&
\alpha2^{n+6} n \bigg(\frac{6\eta^2}{t_{\circ}^2}+\frac{12
\delta\eta}{t}+\frac{3\delta(2\delta
-1)}{t^2}\bigg)^n\\\nonumber&+&\frac{9\beta \mathcal{C}^4
(t_{\circ}\delta +\eta t)^8}{\pi^2 t_{\circ}^4t^4 \bigg(\pi\alpha
t_{\circ}^2t^2+(t_{\circ}\delta +\eta
t)^2\bigg)^2}\\\nonumber&+&\bigg(\alpha t_{\circ}^2\delta2^{n+7}
(n-1) n (t_{\circ}\delta +\eta t) (t_{\circ}(2\delta-1)+2\eta
t)\\\nonumber&\times&\bigg(\frac{6\eta^2}{t_{\circ}^2}+\frac{12\delta\eta
}{t_{\circ}t}+\frac{3\delta(2\delta-1)}{t^2}\bigg)^n
\\\nonumber&\times&\bigg(t_{\circ}^2\delta(2\delta-1)+4t_{\circ}\delta\eta
t+2\eta^2t^2\bigg)^{-2}\bigg)\\\nonumber&+&
\bigg(6\mathcal{C}^2(t_{\circ}\delta +\eta t)^4 \bigg(3\beta
\mathcal{C}^2(t_{\circ}\delta +\eta t)^4-\pi\alpha
t_{\circ}^22^{n+2}nt^2\bigg)
\\\nonumber&\times&\bigg(\frac{6\eta^2}{t_{\circ}^2}+\frac{12\delta\eta}{t_{\circ}t}+
\frac{3\delta(2\delta-1)}{t_{\circ}^2t^2}\bigg)^{n-1}t_{\circ}^2\bigg(\delta^2+\pi\alpha
t^2\bigg)\\\nonumber&+&2\delta\eta t_{\circ}t+\eta^2t^2 +8\pi
t_{\circ}^2t^2\big(\pi\alpha t_{\circ}^2t^2+(t_{\circ}\delta+\eta
t)^2\big)\bigg)
\\\label{33}&\times&\bigg(\pi^2t_{\circ}^4t^4 \big(\pi\alpha t_{\circ}^2t^2+(t_{\circ}\delta+\eta
t)^2\big)^2\bigg)^{-1}\bigg].
\end{eqnarray}
\begin{eqnarray}\nonumber
p_{DE}&=&\frac{6^{-n}}{\alpha
n}\bigg[\frac{2\eta^2}{t_{\circ}^2}+\frac{4\delta\eta}{t_{\circ}
t}+\frac{\delta(2\delta-1)}{t^2}\bigg]^{1-n}\\\nonumber
&\times&\bigg[-\frac{144\alpha\delta(n-1)n(t_{\circ}\delta +\eta t)
(t_{\circ}(2\delta-1)+2\eta t)}{t_{\circ}^2
t^4}\\\nonumber&+&\alpha3^{n+1}\bigg(\frac{4\eta^2}{t_{\circ}^2}+\frac{8
\delta\eta}{t_{\circ}t}+\frac{2\delta(2\delta
-1)}{t^2}\bigg)^n\\\nonumber&-&\alpha3^{n+1}n
\bigg(\frac{4\eta^2}{t_{\circ}^2}+\frac{8\delta\eta}{t_{\circ}t}+\frac{2
\delta(2\delta-1)}{t^2}\bigg)^n\\\nonumber&+&\bigg(\alpha
t_{\circ}^4\delta^22^{n+2}(n-2)(n-1)n(-2t_{\circ}\delta+t_{\circ}-2\eta
t)^2\\\nonumber&\times&\bigg(\frac{6\eta^2}{t_{\circ}^2}+\frac{12\delta\eta
}{t_{\circ}t}+\frac{3\delta(2\delta
-1)}{t^2}\bigg)^n\bigg(t_{\circ}^2\delta(2\delta-1)\\\nonumber&+&4
\delta\eta t_{\circ}t+2\eta^2 t^2\bigg)^{-3}\bigg)+\bigg(\alpha
t_{\circ}^3\delta2^{n+1}(n-1)n\\\nonumber&\times&(t_{\circ}(6\delta
-3)+4\eta t)\bigg(\frac{6\eta^2}{t_{\circ}^2}+\frac{12\delta\eta}{
t_{\circ}t}+\frac{3\delta(2\delta-1)}{t^2}\bigg)^n
\\\nonumber&\times&\bigg(t_{\circ}^2
\delta(2\delta-1)+4t_{\circ}\delta\eta t+2\eta^2
t^2\bigg)^{-2}\bigg)\\\label{34}&-&\frac{27\beta
\mathcal{C}^4(t_{\circ}\delta+\eta t)^8}{64\pi^2 t_{\circ}^4
t^4\bigg(\pi\alpha t_{\circ}^2t^2+(t_{\circ}\delta+\eta
t)^2\bigg)^2}\bigg].
\end{eqnarray}
The EoS parameter for DE turns out to be
\begin{eqnarray}\nonumber
\omega_{DE}&=&\frac{64}{3}\bigg[-\frac{144\alpha\delta(n-1)n(t_{\circ}
\delta +\eta t)(t_{\circ}(2\delta-1)+2\eta t)}{t_{\circ}^2
t^4}\\\nonumber&+&\alpha3^{n+1}\bigg(\frac{4\eta
^2}{t_{\circ}^2}+\frac{8\delta\eta}{
t_{\circ}t}+\frac{2\delta(2\delta
-1)}{t^2}\bigg)^n\\\nonumber&-&\alpha3^{n+1}n\bigg(\frac{4
\eta^2}{t_{\circ}^2}+\frac{8\delta\eta}{t_{\circ}t}+\frac{2\delta(2
\delta -1)}{t^2}\bigg)^n+\bigg(\alpha
t_{\circ}^4\delta^22^{n+2}(n-2)\\\nonumber&\times&
(n-1)n(-2t_{\circ}\delta +t_{\circ}-2\eta
t)^2\bigg(\frac{6\eta^2}{t_{\circ}^2}+\frac{12\delta\eta}{t_{\circ}
t}+\frac{3\delta(2\delta
-1)}{t^2}\bigg)^n\bigg)\\\nonumber&\times&\bigg(t_{\circ}^2\delta(2\delta-1)+4
\delta\eta t_{\circ}t+2\eta^2t^2\bigg)^{-3}+\bigg(\alpha
t_{\circ}^3\delta2^{n+1}(n-1)n\\\nonumber&\times&(t_{\circ}(6\delta-3)+4\eta
t)\bigg(\frac{6\eta^2}{t_{\circ}^2}+\frac{12\delta\eta}{t_{\circ}
t}+\frac{3\delta(2\delta-1)}{t^2}\bigg)^n\bigg)
\\\nonumber&\times&\bigg(t_{\circ}^2\delta(2\delta-1)+4t_{\circ}\delta\eta
t+2\eta^2t^2\bigg)^{-2}\\\nonumber&-&\frac{27\beta
\mathcal{C}^4(t_{\circ} \delta+\eta t)^8}{64\pi^2t_{\circ}^4
t^4\bigg(\pi\alpha t_{\circ}^2t^2+(t_{\circ}\delta+\eta
t)^2\bigg)^2}\bigg]\bigg[\alpha2^{n+6}\bigg(\frac{6\eta
^2}{t_{\circ}^2}\\\nonumber&+&\frac{12\delta\eta}{
t_{\circ}t}+\frac{3\delta(2\delta -1)}{t^2}\bigg)^n-\alpha 2^{n+6} n
\bigg(\frac{6 \eta^2}{t_{\circ}^2}+\frac{12\delta\eta}{t_{\circ}
t}+\frac{3\delta(2\delta-1)}{t^2}\bigg)^n\\\nonumber&+&\frac{9 \beta
\mathcal{C}^4(t_{\circ}\delta +\eta t)^8}{\pi^2 t_{\circ}^4t^4
\bigg(\pi\alpha t_{\circ}^2t^2+(t_{\circ}\delta+\eta
t)^2\bigg)^2}+\bigg(\alpha
t_{\circ}^2\delta2^{n+7}(n-1)n\\\nonumber&\times&(t_{\circ}\delta+\eta
t)(t_{\circ}(2\delta-1)+2\eta
t)\bigg(\frac{6\eta^2}{t_{\circ}^2}+\frac{12\delta\eta}{
t_{\circ}t}+\frac{3\delta(2\delta
-1)}{t^2}\bigg)^n\bigg)\\\nonumber&\times&\bigg(t_{\circ}^2\delta(2\delta-1)+4
\delta\eta t_{\circ}t+2\eta^2
t^2\bigg)^{-2}+\bigg(6\mathcal{C}^2(t_{\circ}\delta +\eta
t)^4\bigg(\pi\alpha
t_{\circ}^2\\\nonumber&\times&\big(-2^{n+2}\big)nt^2
\big(\frac{6\eta^2}{t_{\circ}^2}+\frac{12\delta\eta}{
t_{\circ}t}+\frac{3\delta(2\delta-1)}{t^2}\big)^{n-1}
\big(t_{\circ}^2\big(\delta^2+\pi\alpha
t^2\big)\\\nonumber&+&2t_{\circ}\delta\eta t+\eta^2 t^2\big)+8\pi
t^2_{\circ}t^2 \big(\pi\alpha t_{\circ}^2t^2+(t_{\circ}\delta+\eta
t)^2\big)+3 \beta
\mathcal{C}^2\\\label{35}&\times&(t_{\circ}\delta+\eta
t)^4\big)\bigg)\bigg(\pi^2 t_{\circ}^4t^4\big(\pi\alpha
t_{\circ}^2t^2+(t_{\circ}\delta+\eta
t)^2\big)^2\bigg)^{-1}\bigg]^{-1}.
\end{eqnarray}
Figure \textbf{7} signifies that the energy density for RHDE is
positive, suggesting accelerated cosmic expansion. This accelerated
expansion is verified by the behavior of matter variables for DE, as
energy density is positive and pressure is negative. The EoS
parameter represents the phantom dominated era and stability
analysis ensures a stable accelerating universe.
\begin{figure}
\epsfig{file=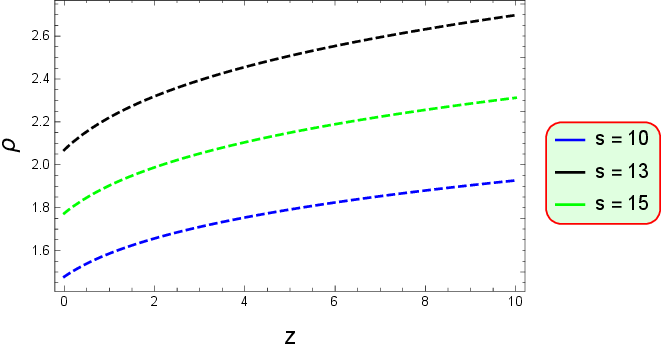,width=.5\linewidth}
\epsfig{file=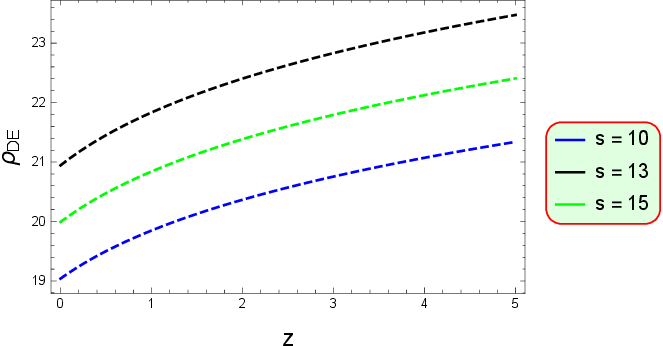,width=.5\linewidth}
\epsfig{file=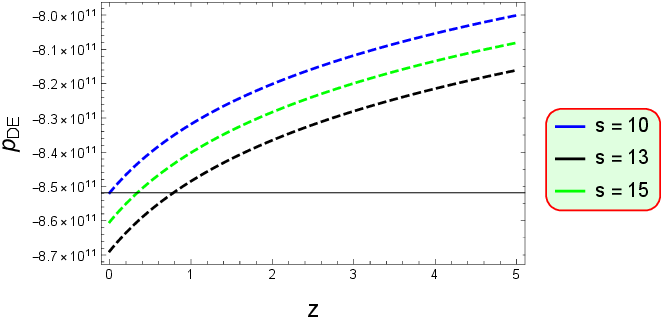,width=.5\linewidth}
\epsfig{file=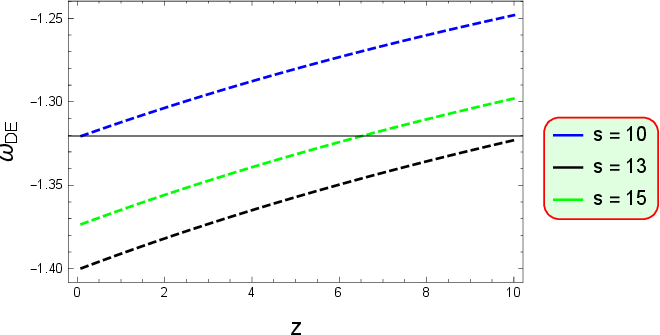,width=.5\linewidth}
\begin{center}
\epsfig{file=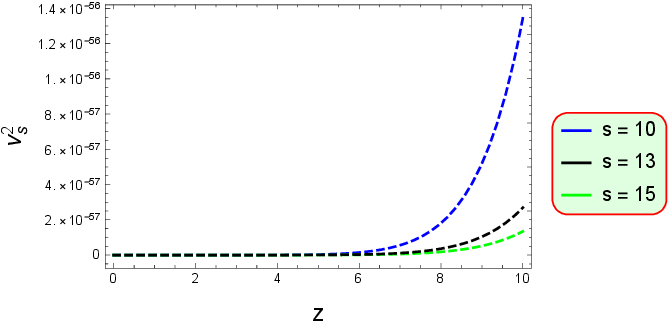,width=.5\linewidth}
\end{center}
\caption{Parametric behavior of $\rho$, $\rho_{DE}$, $p_{DE}$,
$\omega_{DE}$ and $\nu_{s}^2$.}
\end{figure}

\subsection{Sharma-Mittal Holographic Dark Energy Model}

The Sharma-Mittal HDE density associated with the Riici horizon is
derived using Eq. (\ref{26.0}) and the expression for the Ricci
horizon as
\begin{equation}\label{36}
\rho_{SMDE}=\frac{3 \mathcal{C}^2}{\mathcal{B}}
\bigg(\frac{\delta\eta}{t_{\circ}}+\frac{\delta (\delta t-\beta
)}{t^2}\bigg)^2 \bigg(\frac{\pi\chi
}{\frac{\delta\eta}{t_{\circ}}+\frac{\delta(\delta t-\beta
)}{t^2}}+1\bigg)^{\mathcal{B}\chi}-1.
\end{equation}
Using Eqs.(\ref{18}), (\ref{19}) and (\ref{36}) into Eqs.(\ref{11})
and (\ref{12}), the expression for energy density and pressure for
DE turn out to be
\begin{eqnarray}\nonumber
\rho_{DE}&=&\frac{2^{-n}3^{1-n}}{\alpha n}
\bigg(\frac{2\eta^2}{\mathcal{B}^2}+\frac{4\delta\eta}{\mathcal{B}
t}+\frac{\delta(2\delta -1)}{t^2}\bigg)^{1-n}
\\\nonumber&\times&\bigg[\alpha6^n\bigg(\frac{2\eta
^2}{\mathcal{B}^2}+\frac{4\delta\eta}{\mathcal{B}t}+\frac{\delta (2
\delta -1)}{t^2}\bigg)^n\\\nonumber&-&\alpha 6^nn
\bigg(\frac{2\eta^2}{\mathcal{B}^2}+\frac{4\delta\eta}{\mathcal{B}
t}+\frac{\delta(2\delta-1)}{t^2}\bigg)^n\\\nonumber&+& \frac{\beta
\bigg( t_{\circ}^2\mathcal{B} t^4-3 \mathcal{C}^2\delta^2
\bigg(-t_{\circ}\beta+t_{\circ}\delta t+\eta t^2\bigg)^2
\bigg(\frac{\pi\chi
}{\frac{\delta\eta}{t_{\circ}}+\frac{\delta(\delta
t-\beta)}{t^2}}+1\bigg)^{\mathcal{B}\chi }\bigg)^2}{t_{\circ}^4
\mathcal{B}^2 t^8}\\\nonumber&+& \bigg(\alpha  t_{\circ}^2 \delta
2^{n+1}(n-1)n(t_{\circ}\delta+\eta t) (t_{\circ}(2\delta-1)+2\eta
t)\\\nonumber&\times&\bigg(\frac{6\eta^2}{t_{\circ}^2}+\frac{12\delta\eta
}{t_{\circ}
t}+\frac{3\delta(2\delta-1)}{t^2}\bigg)^n\bigg)\\\nonumber&\times&\bigg(t_{\circ}^2
\delta(2\delta-1)+4 b\delta\eta t+2\eta^2
t^2\bigg)^{-2}\\\nonumber&-&\bigg(t_{\circ}^2\mathcal{B}t^4-3
\mathcal{C}^2 \delta^2 \bigg(-t_{\circ}\beta+t_{\circ}\delta t+\eta
t^2\bigg)^2\\\nonumber&\times& \bigg(\frac{\pi t_{\circ} t^2 \chi
}{\delta \bigg(-t_{\circ}\beta+t_{\circ}\delta t+\eta
t^2\bigg)}+1\bigg)^{\mathcal{B}\chi}
\\\nonumber&\times&\bigg(t_{\circ}^2 \mathcal{B}t^4-\alpha t_{\circ}^2
\mathcal{B} 6^{n-1} nt^4 \bigg(\frac{2\eta^2}{t_{\circ}^2} +\frac{4
\delta\eta }{ t_{\circ}t}
+\frac{\delta(2\delta-1)}{t^2}\bigg)^{n-1}\\\nonumber&-&
\beta\bigg(\mathcal{B}t_{\circ}^2 t^4-3 \mathcal{C}^2 \delta^2
\bigg(-t_{\circ} \beta+t_{\circ}\delta t+\eta t^2\bigg)^2
\\\label{37}&\times& \bigg(\frac{\pi t_{\circ} t^2\chi}{\delta
\bigg(-t_{\circ}\beta+t_{\circ}\delta t+\eta
t^2\bigg)}+1\bigg)^{\mathcal{B}\chi } \bigg)\bigg)\bigg(t_{\circ}^4
\mathcal{B}^2 t^8\bigg)^{-1}\bigg].
\end{eqnarray}
\begin{eqnarray}\nonumber
p_{DE}&=&\frac{6^{-n}\bigg(\frac{2\eta^2}{t_{\circ}^2}+\frac{4\delta\eta}{t_{\circ}
t}+\frac{\delta(2\delta-1)}{t^2}\bigg)^{1-n}}{\alpha
n}\\\nonumber&\times&\bigg[-\frac{144\alpha\delta(n-1)n(t_{\circ}\delta
+\eta t)(t_{\circ}(2\delta-1)+2\eta t)}{t_{\circ}^2 t^4}
\\\nonumber&+&\bigg(\alpha t_{\circ}^4\delta^22^{n+2}
(n-2)(n-1)n(b(2\delta-1)+2\eta
t)^2\\\nonumber&\times&\bigg(\frac{6\eta
^2}{t_{\circ}^2}+\frac{12\delta\eta}{
t_{\circ}t}+\frac{3\delta(2\delta -1)}{t^2}\bigg)^n\bigg)
\\\nonumber&\times&\bigg(t_{\circ}^2\delta(2\delta-1)
+4t_{\circ}\delta\eta t+2\eta^2
t^2\bigg)^{-3}\\\nonumber&+&\bigg(\alpha
t_{\circ}^3\delta2^{n+1}(n-1)n(t_{\circ}(6 \delta-3)+4\eta
t)\\\nonumber&\times&\bigg(\frac{6\eta
^2}{t_{\circ}^2}+\frac{12\delta\eta}{t_{\circ}
t}+\frac{3\delta(2\delta
-1)}{t^2}\bigg)^n\bigg)\\\nonumber&\times&\bigg(t_{\circ}^2\delta(2\delta-1+4
t_{\circ}\delta\eta
t+2 \eta^2 t^2\bigg)^{-2}\\\nonumber&+&\frac{3}{t_{\circ}^4
\mathcal{B}^2 t^8}\bigg(\alpha b^4 \mathcal{B}^26^n t^8
\bigg(\frac{2\eta^2}{t_{\circ}^2}+\frac{4\delta\eta}{ t_{\circ}t}
+\frac{\delta(2\delta-1)}{t_{\circ}^2t^2}\bigg)^n
\\\nonumber&-&\alpha t_{\circ}^4 \mathcal{B}^26^n
nt^8\bigg(\frac{2\eta
^2}{t_{\circ}^2}+\frac{4\delta\eta}{t_{\circ}t}+\frac{\delta(2\delta
-1)}{t^2}\bigg)^n\\\nonumber&-&\beta\bigg(t_{\circ}^2 \mathcal{B}
t^4-3 \mathcal{C}^2\delta^2 \bigg(-t_{\circ}\beta+t_{\circ}\delta
t_{\circ}t+\eta
t_{\circ}^2t^2\bigg)^2\\\label{38}&\times&\bigg(\frac{\pi\chi
}{\frac{\delta\eta}{t_{\circ}}+\frac{\delta(\delta t-\beta
)}{t^2}}+1\bigg)^{\mathcal{B}\chi }\bigg)^{2}\bigg)\bigg)\bigg]
\end{eqnarray}
The EoS parameter for SMHDE model associated with Ricci horizon is
given as
\begin{eqnarray}\nonumber
\omega_{DE}&=&\bigg[-\frac{144\alpha\delta(n-1)n(t_{\circ}\delta
+\eta t)(t_{\circ}(2\delta-1)+2\eta t)}{t_{\circ}^2 t^4}
\\\nonumber&+&\bigg(\alpha t_{\circ}^4\delta^22^{n+2}
(n-2)(n-1)n(t_{\circ}(2\delta-1)+2\eta
t)^2\\\nonumber&\times&\bigg(\frac{6\eta
^2}{t_{\circ}^2}+\frac{12\delta\eta}{
t_{\circ}t}+\frac{3\delta(2\delta -1)}{t^2}\bigg)^n\bigg)
\\\nonumber&\times&\bigg(t_{\circ}^2\delta(2\delta-1)
+4\delta\eta tt_{\circ}+2\eta^2
t^2\bigg)^{-3}\\\nonumber&+&\bigg(\alpha
t_{\circ}^3\delta2^{n+1}(n-1)n(t_{\circ}(6 \delta-3)+4\eta
t)\\\nonumber&\times&\bigg(\frac{6\eta
^2}{t_{\circ}^2}+\frac{12\delta\eta}{t_{\circ}
t}+\frac{3\delta(2\delta
-1)}{t^2}\bigg)^n\bigg)\\\nonumber&\times&\bigg(t_{\circ}^2\delta(2\delta-1+4t_{\circ}\delta\eta
t+2 \eta^2 t^2\bigg)^{-2}\\\nonumber&+&\frac{3}{t_{\circ}^4
\mathcal{B}^2 t^8}\bigg(\alpha t_{\circ}^4 \mathcal{B}^26^n t^8
\bigg(\frac{2\eta^2}{t_{\circ}^2}+\frac{4\delta\eta}{ t_{\circ}t}
+\frac{\delta(2\delta-1)}{t^2}\bigg)^n
\\\nonumber&-&\alpha t_{\circ}^4 \mathcal{B}^26^n
nt^8\bigg(\frac{2\eta
^2}{t_{\circ}^2}+\frac{4\delta\eta}{t_{\circ}t}+\frac{\delta(2\delta
-1)}{t^2}\bigg)^n\\\nonumber&-&\beta\bigg(t_{\circ}^2 \mathcal{B}
t^4-3 \mathcal{C}^2\delta^2 \bigg(-t_{\circ}\beta+t_{\circ}\delta
t+\eta t^2\bigg)^2\\\nonumber&\times&\bigg(\frac{\pi\chi
}{\frac{\delta\eta}{t_{\circ}}+\frac{\delta(\delta t-\beta
)}{t^2}}+1\bigg)^{\mathcal{B}\chi}\bigg)^{2}\bigg)\bigg)\bigg]
\\\nonumber&\times&
\bigg[\alpha6^n\bigg(\frac{2\eta
^2}{\mathcal{B}^2}+\frac{4\delta\eta}{\mathcal{B}t}+\frac{\delta (2
\delta -1)}{t^2}\bigg)^n\\\nonumber&-&\alpha 6^nn
\bigg(\frac{2\eta^2}{\mathcal{B}^2}+\frac{4\delta\eta}{\mathcal{B}
t}+\frac{\delta(2\delta-1)}{t^2}\bigg)^n\\\nonumber&+& \frac{\beta
\bigg( t_{\circ}^2\mathcal{B} t^4-3 \mathcal{C}^2\delta^2
\bigg(-t_{\circ}\beta+t_{\circ}\delta t+\eta t^2\bigg)^2
\bigg(\frac{\pi\chi
}{\frac{\delta\eta}{t_{\circ}}+\frac{\delta(\delta
t-\beta)}{t^2}}+1\bigg)^{\mathcal{B}\chi }\bigg)^2}{t_{\circ}^4
\mathcal{B}^2 t^8}\\\nonumber&+& \bigg(\alpha  t_{\circ}^2 \delta
2^{n+1}(n-1)n(t_{\circ}\delta+\eta t) (t_{\circ}(2\delta-1)+2\eta
t)\\\nonumber&\times&\bigg(\frac{6\eta^2}{t_{\circ}^2}+\frac{12\delta\eta
}{t_{\circ}
t}+\frac{3\delta(2\delta-1)}{t^2}\bigg)^n\bigg)\\\nonumber&\times&\bigg(t_{\circ}^2
\delta(2\delta-1)+4 t_{\circ}\delta\eta t+2\eta^2
t^2\bigg)^{-2}\\\nonumber&-&\bigg(t_{\circ}^2\mathcal{B}t^4-3
\mathcal{C}^2 \delta^2 \bigg(-t_{\circ}\beta+t_{\circ}\delta t+\eta
t^2\bigg)^2\\\nonumber&\times& \bigg(\frac{\pi t_{\circ} t^2 \chi
}{\delta \bigg(-t_{\circ}\beta+t_{\circ}\delta t+\eta
t^2\bigg)}+1\bigg)^{\mathcal{B}\chi}
\\\nonumber&\times&\bigg(t_{\circ}^2 \mathcal{B}t^4-\alpha t_{\circ}^2
\mathcal{B} 6^{n-1} nt^4 \bigg(\frac{2\eta^2}{t_{\circ}^2} +\frac{4
\delta\eta }{t_{\circ}t}
+\frac{\delta(2\delta-1)}{t^2}\bigg)^{n-1}\\\nonumber&-&
\beta\bigg(t_{\circ}^2 \mathcal{B} t^4-3 \mathcal{C}^2 \delta^2
\bigg(-t_{\circ} beta+t_{\circ}\delta t+\eta t^2\bigg)^2
\\\label{39}&\times& \bigg(\frac{\pi t_{\circ} t^2\chi}{\delta
\bigg(-t_{\circ}\beta+t_{\circ}\delta t+\eta
t^2\bigg)}+1\bigg)^{\mathcal{B}\chi } \bigg)\bigg)\bigg(t_{\circ}^4
\mathcal{B}^2 t^8\bigg)^{-1}\bigg]^{-1}.
\end{eqnarray}
The graphical analysis presented in Figure \textbf{8} determines an
accelerated expansion of the universe, characterized by a positive
energy density and behavior of matter variables for DE. The
trajectories of the EoS parameter indicate a quintessence era of DE
and the squared sound speed parameter confirms the stability of the
model.
\begin{figure}
\epsfig{file=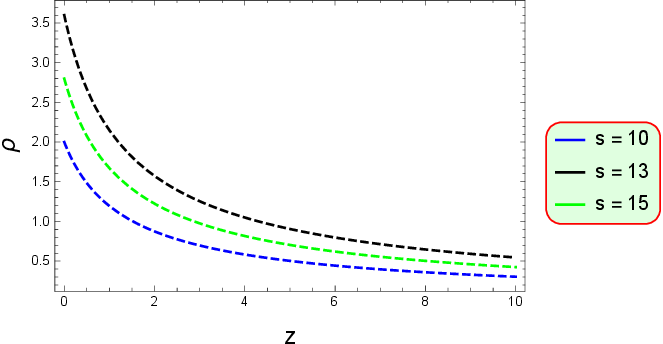,width=.5\linewidth}
\epsfig{file=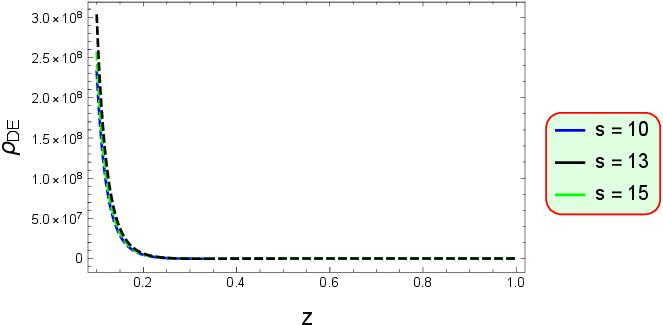,width=.5\linewidth}
\epsfig{file=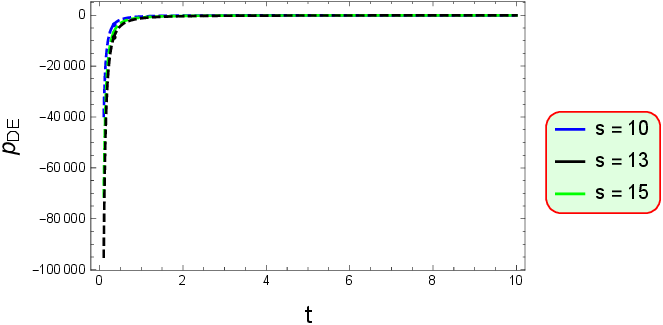,width=.5\linewidth}
\epsfig{file=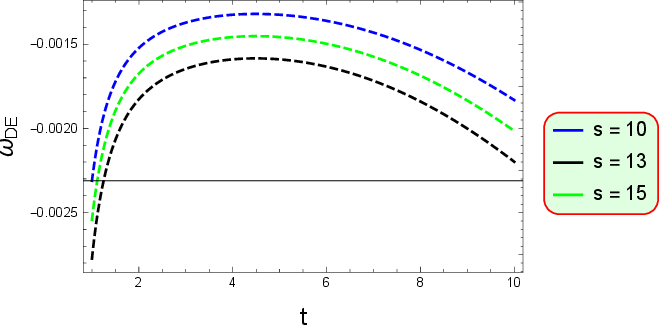,width=.5\linewidth}
\begin{center}
\epsfig{file=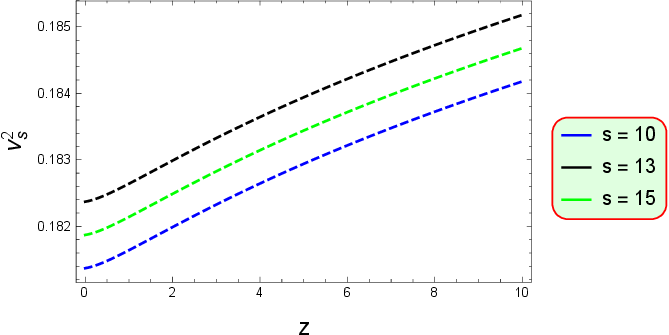,width=.5\linewidth}
\end{center}
\caption{Parametric evolution of $\rho$, $\rho_{DE}$, $p_{DE}$,
$\omega_{DE}$ and $\nu_{s}^2$.}
\end{figure}

\subsection{Generalized Holographic Dark Energy Model}

The expression for the GHDE for the Ricci horizon is given as
\begin{equation}\label{40}
\rho_{GHDE}=\frac{3\mathcal{C}^2}{8\pi \mathcal{B}}
\bigg(\frac{\delta \eta }{t_{\circ}}+\frac{\delta (\delta t-\beta
)}{t^2}\bigg) \bigg(\bigg(\frac{\pi
\alpha}{\lambda\bigg(\frac{\delta\eta
}{t_{\circ}}+\frac{\delta(\delta t-\beta
)}{t^2}\bigg)}+1\bigg)^{\lambda }-1\bigg).
\end{equation}
The energy density and pressure for this DE model are obtained using
Eq.(\ref{40}) and Eq.(\ref{19}) into Eqs.(\ref{11}) and (\ref{12})
as
\begin{eqnarray}\nonumber
\rho_{DE}&=&\frac{2^{-n-6}3^{1-n}\bigg(\frac{2\eta^2}{t_{\circ}^2}+\frac{4
\delta\eta}{t_{\circ}t}+\frac{\delta(2\delta
-1)}{t^2}\bigg)^{1-n}}{\pi^2\alpha t_{\circ}^2 \mathcal{B}^2 n
t^4}\\\nonumber&\times&\bigg[\bigg(\pi^2 \alpha t_{\circ}^4
\mathcal{B}^2\delta 2^{n+7} (n-1) n t^4(t_{\circ}\delta +\eta
t_{\circ}) (t_{\circ}(2\delta-1)\\\nonumber&+&2\eta
t)\bigg(\frac{6\eta^2}{t_{\circ}^2}+\frac{12\delta\eta}{t_{\circ}
t}+\frac{3\delta(2\delta
-1)}{t^2}\bigg)^n\bigg)\\\nonumber&\times&\bigg(t_{\circ}^2 \delta
(2\delta-1)+4 t_{\circ}\delta\eta t+2\eta^2
t^2\bigg)^{-2}\\\nonumber&+&t_{\circ}^2\bigg(\pi^2\alpha
\mathcal{B}^2 (-2^{n+6}) (n-1)t^4
\bigg(\frac{6\eta^2}{t_{\circ}^2}+\frac{12\delta\eta}{t_{\circ}
t}\\\nonumber&+&\frac{3\delta(2\delta
-1)}{t^2}\bigg)^n+9\beta\mathcal{C}^4\delta^2 (\beta-\delta
t)^2\bigg(\bigg(\frac{\pi\alpha}{\lambda
\bigg(\frac{\delta\eta}{t_{\circ}}+\frac{\delta(\delta t-\beta
)}{t^2}\bigg)}\\\nonumber&+&1\bigg)^{\lambda}-1\bigg)^2\bigg)+6
\mathcal{C}^2 \delta(t_{\circ}(\beta-\delta t)-\eta t^2)+6
\mathcal{C}^2 \delta(t_{\circ}(\beta-\delta t)-\eta
t^2)\\\nonumber&\times& 8\pi b \mathcal{B}t^2-3 \mathcal{C}^2\beta
\delta (t_{\circ}(\beta -\delta t)-\eta t^2)\bigg(\bigg(\frac{\pi
\alpha}{\lambda\bigg(\frac{\delta\eta}{t_{\circ}}+\frac{\delta(\delta
t-\beta)}{t^2}\bigg)}\\\nonumber&+&1\bigg)^{\lambda
}-1\bigg)\bigg(1-\bigg(\frac{\pi\alpha}{\lambda
\bigg(\frac{\delta\eta }{t_{\circ}}+\frac{\delta(\delta t-\beta
)}{t^2}\bigg)}+1\bigg)^{\lambda }\\\nonumber&-&\frac{\pi\alpha
t_{\circ}^3 \mathcal{B} 2^{n+2} 3^{n-1} n t^4 \bigg(\frac{2 \eta
^2}{t_{\circ}^2}+\frac{4\delta\eta}{t_{\circ}
t}+\frac{\delta(2\delta
-1)}{t^2}\bigg)^n}{t_{\circ}^2\delta(2\delta-1)+4
t_{\circ}\delta\eta t+2 \eta ^2 t^2}\bigg)\\\nonumber&+&18
t_{\circ}\beta \mathcal{C}^4 \delta^2 \eta t^2 (\delta t-\beta
)\bigg(\bigg(\frac{\pi\alpha }{\lambda
\bigg(\frac{\delta\eta}{t_{\circ}}+\frac{\delta (\delta t-\beta
)}{t^2}\bigg)}+1\bigg)^{\lambda
}-1\bigg)^2\\\label{41}&+&9\beta\mathcal{C}^4\delta^2\eta^2 t^4
\bigg(\bigg(\frac{\pi\alpha}{\lambda
\bigg(\frac{\delta\eta}{t_{\circ}}+\frac{\delta(\delta t-\beta
)}{t^2}\bigg)}+1\bigg)^{\lambda }-1\bigg)^2\bigg].
\end{eqnarray}
\begin{eqnarray}\nonumber
p_{DE}&=&\frac{2^{-n-6}3^{-n}\bigg(\frac{2\eta^2}{t_{\circ}^2}+\frac{4
\delta\eta}{t_{\circ}t}+\frac{\delta(2\delta
-1)}{t^2}\bigg)^{1-n}}{\alpha t_{\circ}^2 n}\\\nonumber&\times&
\bigg[-\frac{9216\alpha\delta(n-1)n(t_{\circ}\delta +\eta
t)(t_{\circ}(2 \delta-1)+2\eta t)}{t^4}\\\nonumber&+&18 t_{\circ}
\beta \mathcal{C}^4 \delta^2\eta t^2(\delta t-\beta)
\bigg(\bigg(\frac{\pi\alpha}{\lambda\bigg(\frac{\delta\eta
}{t_{\circ}}+\frac{\delta(\delta t-\beta
)}{t^2}\bigg)}+1\bigg)^{\lambda }-1\bigg)^2\\\nonumber&+&9\beta
\mathcal{C}^4 \delta^2 \eta^2 t^4
\bigg(\bigg(\frac{\pi\alpha}{\lambda
\bigg(\frac{\delta\eta}{t_{\circ}}+\frac{\delta(\delta t-\beta
)}{t^2}\bigg)}+1\bigg)^{\lambda}-1\bigg)^2\\\nonumber&+&9 \beta
\mathcal{C}^4\delta^2 (\beta-\delta t)^2 \bigg(\bigg(\frac{\pi
\alpha}{\lambda\bigg(\frac{\delta\eta}{t_{\circ}}+\frac{\delta(\delta
t-\beta)}{t^2}\bigg)}+1\bigg)^{\lambda
}-1\bigg)^2\bigg)\\\nonumber&-&\frac{1}{\pi^2 \mathcal{B}^2
t^4}3\bigg(t_{\circ}^2\bigg(\pi^2 \alpha\mathcal{B}^22^{n+6}(n-1)
t^4 \bigg(\frac{6\eta^2}{t_{\circ}^2}+\frac{12\delta\eta}{t_{\circ}
t}\\\nonumber&+&\frac{3\delta(2\delta-1)}{t^2}\bigg)^n\bigg)\bigg)+\bigg(\alpha
t_{\circ}^6\delta^2
2^{n+8}(n-2)(n-1)n(t_{\circ}(2\delta-1)\\\nonumber&+&2\eta
t)^2\bigg(\frac{6\eta^2}{t_{\circ}^2}+\frac{12\delta\eta}{t_{\circ}
t}+\frac{3\delta(2\delta-1)}{t^2}\bigg)^n\bigg(t_{\circ}^2 \delta (2
\delta -1)+4 t_{\circ} \delta  \eta t\\\nonumber&+&2\eta^2
t^2\bigg)^{-3}\bigg)+\bigg(\alpha
t_{\circ}^5\delta2^{n+7}(n-1)n(t_{\circ}(6\delta-3)+4\eta
t)\\\nonumber&\times&\bigg(\frac{6\eta
^2}{t_{\circ}^2}+\frac{12\delta\eta}{t_{\circ}
t}+\frac{3\delta(2\delta
-1)}{t^2}\bigg)^n\bigg)\bigg(t_{\circ}^2\delta (2
\delta-1)+4t_{\circ} \delta\eta t\\\label{42}&+&2\eta^2
t^2\bigg)^{-2}\bigg)\bigg].
\end{eqnarray}
The EoS parameter turns out to be
\begin{eqnarray}\nonumber
\omega_{DE}&=&\bigg[\pi^2 \mathcal{B}^2
t^4\bigg(-\frac{9216\alpha\delta(n-1)n(t_{\circ}\delta +\eta t)
(t_{\circ}(2\delta-1)+2\eta t)}{t^4}\\\nonumber&+&\bigg(\alpha
t_{\circ}^6\delta^2 2^{n+8}(n-2)(n-1)n(t_{\circ}(2\delta-1)+2\eta
t)^2\\\nonumber&\times&\bigg(\frac{6\eta^2}{t_{\circ}^2}+\frac{12\delta\eta
}{t_{\circ} t}+\frac{3\delta(2\delta-1)}{t^2}\bigg)^n
\bigg(t_{\circ}^2\delta(2 \delta-1)+4 t_{\circ}\delta\eta
t\\\nonumber&+&2\eta^2 t^2\bigg)^{-3}+\bigg(\alpha t_{\circ}^5\delta
2^{n+7}(n-1)n(t_{\circ}(6\delta-3)+4 \eta
t)\\\nonumber&\times&\bigg(\frac{6\eta^2}{t_{\circ}^2}+\frac{12\delta\eta}{t_{\circ}
t}+\frac{3\delta(2\delta
-1)}{t^2}\bigg)^n\bigg)\bigg(t_{\circ}^2\delta(2\delta-1)+4
t_{\circ} \delta\eta t\\\nonumber&+&2\eta^2
t^2\bigg)^{-2}\bigg)-\frac{3}{\pi^2 \mathcal{B}^2
t^4}\bigg(t_{\circ}^2\bigg(\pi^2 \alpha\mathcal{B}^22^{n+6}(n-1) t^4
\bigg(\frac{6\eta^2}{t_{\circ}^2}+\frac{12\delta\eta}{t_{\circ}
t}\\\nonumber&+&\frac{3\delta(2\delta-1)}{t^2}\bigg)^n\bigg)\bigg)+\bigg(\alpha
t_{\circ}^6\delta^2
2^{n+8}(n-2)(n-1)n(t_{\circ}(2\delta-1)\\\nonumber&+&2\eta
t)^2\bigg(\frac{6\eta^2}{t_{\circ}^2}+\frac{12\delta\eta}{t_{\circ}
t}+\frac{3\delta(2\delta-1)}{t^2}\bigg)^n\bigg(t_{\circ}^2 \delta (2
\delta -1)+4 t_{\circ} \delta \eta t\\\nonumber&+&2\eta^2
t^2\bigg)^{-3}\bigg)+18t_{\circ}\beta \mathcal{C}^4 \delta^2\eta
t^2(\delta t-\beta)
\bigg(\bigg(\frac{\pi\alpha}{\lambda\bigg(\frac{\delta\eta
}{t_{\circ}}+\frac{\delta(\delta t-\beta
)}{t^2}\bigg)}+1\bigg)^{\lambda }-1\bigg)^2\\\nonumber&+&9\beta
\mathcal{C}^4 \delta^2 \eta^2 t^4
\bigg(\bigg(\frac{\pi\alpha}{\lambda
\bigg(\frac{\delta\eta}{t_{\circ}}+\frac{\delta(\delta t-\beta
)}{t^2}\bigg)}+1\bigg)^{\lambda}-1\bigg)^2\bigg)\bigg]
\\\nonumber&\times&\bigg[3\bigg(\bigg(\pi^2 \alpha t_{\circ}^4
\mathcal{B}^2\delta 2^{n+7} (n-1) nt^4(t_{\circ}\delta +\eta t)
(t_{\circ} (2\delta-1)\\\nonumber&+&2\eta t)\bigg(\frac{6\eta
^2}{t_{\circ}^2}+\frac{12\delta\eta }{t_{\circ}
t}+\frac{3\delta(2\delta
-1)}{t^2}\bigg)^n\bigg)\\\nonumber&\times&\bigg(t_{\circ}^2 \delta
(2 \delta-1)+4 t_{\circ} \delta \eta  t+2 \eta ^2
t^2\bigg)^{-2}\\\nonumber&+&t_{\circ}^2\bigg(\pi^2\alpha
\mathcal{B}^2 (-2^{n+6}) (n-1)t^4
\bigg(\frac{6\eta^2}{t_{\circ}^2}+\frac{12 \delta\eta}{t_{\circ}
t}\\\nonumber&+&\frac{3\delta(2\delta
-1)}{t^2}\bigg)^n+9\beta\mathcal{C}^4\delta^2 (\beta-\delta
t)^2\bigg(\bigg(\frac{\pi\alpha}{\lambda
\bigg(\frac{\delta\eta}{t_{\circ}}+\frac{\delta(\delta t-\beta
)}{t^2}\bigg)}\\\nonumber&+&1\bigg)^{\lambda}-1\bigg)^2\bigg)+6
\mathcal{C}^2 \delta(t_{\circ}(\beta-\delta t)-\eta t^2)+6
\mathcal{C}^2 \delta(t_{\circ}(\beta-\delta t)-\eta
t^2)\\\nonumber&\times& 8\pi t_{\circ} \mathcal{B} t^2-3
\mathcal{C}^2\beta \delta (t_{\circ}(\beta -\delta t)-\eta
t^2)\bigg(\bigg(\frac{\pi \alpha }{\lambda
\bigg(\frac{\delta\eta}{t_{\circ}}+\frac{\delta(\delta t-\beta
)}{t^2}\bigg)}\\\nonumber&+&1\bigg)^{\lambda
}-1\bigg)\bigg(1-\bigg(\frac{\pi\alpha}{\lambda
\bigg(\frac{\delta\eta }{t_{\circ}}+\frac{\delta(\delta t-\beta
)}{t^2}\bigg)}+1\bigg)^{\lambda }\\\nonumber&-&\frac{\pi\alpha
t_{\circ}^3 \mathcal{B} 2^{n+2} 3^{n-1} n t^4 \bigg(\frac{2 \eta
^2}{t_{\circ}^2}+\frac{4 \delta\eta}{t_{\circ}
t}+\frac{\delta(2\delta
-1)}{t^2}\bigg)^n}{t_{\circ}^2\delta(2\delta-1)+4
t_{\circ}\delta\eta t+2 \eta ^2 t^2}\bigg)\\\nonumber&+&18
t_{\circ}\beta \mathcal{C}^4 \delta^2 \eta t^2 (\delta t-\beta
)\bigg(\bigg(\frac{\pi\alpha }{\lambda
\bigg(\frac{\delta\eta}{t_{\circ}}+\frac{\delta (\delta t-\beta
)}{t^2}\bigg)}+1\bigg)^{\lambda
}-1\bigg)^2\\\label{43}&+&9\beta\mathcal{C}^4\delta^2\eta^2 t^4
\bigg(\bigg(\frac{\pi\alpha}{\lambda
\bigg(\frac{\delta\eta}{t_{\circ}}+\frac{\delta(\delta t-\beta
)}{t^2}\bigg)}+1\bigg)^{\lambda }-1\bigg)^2\bigg)\bigg]^{-1}.
\end{eqnarray}
Figure \textbf{9} demonstrates the cosmological evolution of the
energy density, pressure, EoS parameter and squared sound speed
parameter, which collectively describe the stable cosmic accelerated
expansion.
\begin{figure}
\epsfig{file=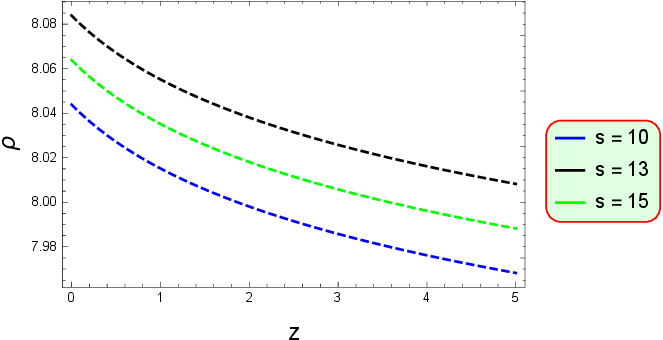,width=.5\linewidth}
\epsfig{file=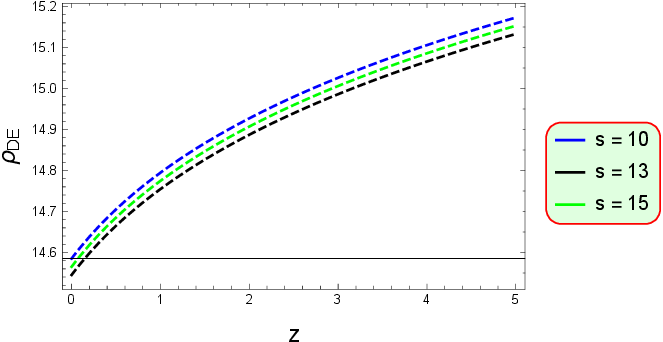,width=.5\linewidth}
\epsfig{file=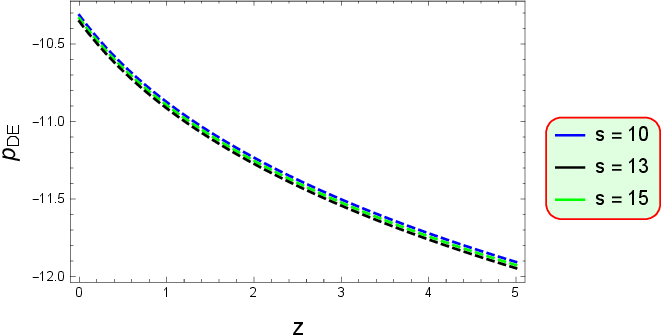,width=.5\linewidth}
\epsfig{file=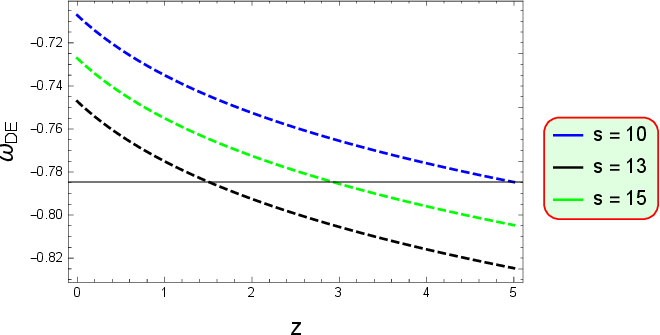,width=.5\linewidth}
\begin{center}
\epsfig{file=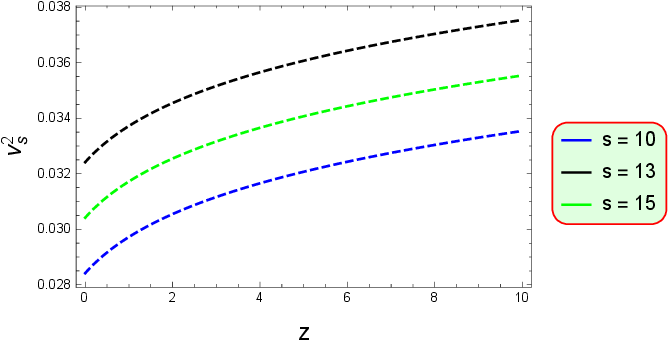,width=.5\linewidth}
\end{center}
\caption{Parametric behavior of $\rho$, $\rho_{DE}$, $p_{DE}$,
$\omega_{DE}$ and $\nu_{s}^2$.}
\end{figure}

\section{Conclusions}

Modified gravity theories have emerged as essential frameworks for
addressing the limitations of GR. Among these, $f(R,T^2)$ gravity
provides a novel extension by incorporating nonlinear contributions
of the trace of the energy-momentum tensor. This extension allows
for a more refined description of matter-energy interactions,
offering deep insights into cosmic evolution. By incorporating self
contraction of stress energy tensor $T^2$, this theory gives a
significant understanding of how matter fields effects gravitational
dynamics. This inclusion plays a comprehensive role in studying the
complexities of energy exchanges between DE, dark matter and
ordinary matter. Consequently, the study of EMSG is pivotal in
exploring our understanding of fundamental relations between matter
and gravity.

In this manuscript, we have studied the cosmic dynamics by exploring
the RHDE, SHMDE and GHDE models for Hubble horizon and Ricci horizon
as IR cutoff in the background of EMSG. To achieve a deep
understanding of cosmic evolution, we have calculated key
cosmological parameters including scale factor, Hubble parameter,
deceleration parameter and EoS parameter. By exploring these
parameters along with the matter variables we have demonstrated how
these variants of HDE models explain the cosmic accelerated
expansion. Our study gives a comprehensive understanding of the
cosmic evolution, offering important findings on the behavior of DE
and its role on the cosmic accelerated expansion. The key finding of
our analysis as follows.

The scale factor and Hubble parameter show positive values, this
refers accelerated cosmic expansion. The negative deceleration
parameter confirms the accelerated expansions which occur at later
times. The Hubble parameter as a function of redshift $H(z)$ agrees
with the Planck observation data and mimics geometric DE, without
the need of cosmological constant in $\Lambda$CDM, this overcomes
the fine-tuning problem \cite{34a} (Figure \textbf{1}). There are
slight discrepancies at the redshifts  where $(1 < z < 2)$. This
reflects the influence of $T^2$ and is consistent with DESI or
Euclid. The energy density remains positive which supports role of
the variant of HDE models in the cosmic accelerated expansion. The
HDE models show adaptability for the various phases in cosmic
acceleration. The EoS parameter significantly determines the phantom
or quintessence behavior, illustrating the model adaptability in
capturing different phases of cosmic acceleration. These variants of
HDE models were found to be stable against perturbations through the
squared sound speed parameter, ensuring its physical viability.
These findings collectively suggest that these DE models are
compelling candidate for examining the cosmic late-time accelerated
expansion in EMSG.

Jawad et al \cite{33} studied the cosmic accelerated expansion
through the HDE models, their investigation did not perform the
stability analysis of these models. However, our study gives a more
significant analysis by determining that these DE models not only
explain comsic acceleration but also show stability through the
squared sound speed parameter. Moreover, we determines a clear
understanding of the EoS parameter, which transition from the
phantom regime in the early cosmos to the quintessence era at late
times, addressing the model's agreement with observational data.
Furthermore, our results shows consistency with the work of Saleem
and Ijaz \cite{34}, validating the robustness of our approach. By
providing a more refined analysis of cosmic dynamics, including
stability and the evolving nature of DE, our research offers an
enhanced understanding of the universe accelerated expansion in
the framework of $f(R,T^2)$ gravity.\\\\
\textbf{Data Availability Statement:} No new data were generated or
analyzed in support of this research.\\\\
\textbf{Credit Author Statement:} \textbf{M. Sharif:}
Conceptualization, Supervision, Writing- Reviewing and Editing.
\textbf{M. Zeeshan Gul:} Investigation, Formal analysis.
\textbf{Imran Hashim:} Methodology, Software, Writing- Original
draft preparation.


\begin{thebibliography}{55}

\bibitem{5}  Buchdahl, H. A.: Mon. Not. R. Astron. Soc.
\textbf{150}(1970)1.

\bibitem{6} Dolgov, A.D. and Kawasaki, M.: Phys. Lett. B \textbf{573}(2003)1;
Capozziello, S., Cardone, V.F. and Troisi, A.: Phys. Rev. D
\textbf{71}(2005)043503.

\bibitem{z1} Adeel, M. et al.: Mod. Phys. Lett. A \textbf{38}(2023)2350152.

\bibitem{z2} Rani, S. et al.: Int. J. Geom. Methods Mod. Phys.
\textbf{21}(2024)2450033.

\bibitem{z3} Waseem, A. et al.: Eur. Phys. J. C
\textbf{83}(2023)1088.

\bibitem{6aa} Mustafa, G. et al.: Eur. Phys. J. C
\textbf{84}(2023)690; Ann. Phys. \textbf{460}(2024)169551; Chin. J.
Phys. \textbf{88}(2024)32

\bibitem{6ddd} Asad, H. et al.: Phys. Dark Universe \textbf{46}(2024)
101666.

\bibitem{z4} Nan G. et al.: Phys. Dark Universe \textbf{46}(2024)101635.

\bibitem{6g} Yousaf, M. Asad, H.: Phys. Dark Universe
\textbf{48}(2025)101841; Yousaf, M.: Chin. J. Phys.
\textbf{95}(2025)1278.

\bibitem{6gg} Dai, E. et al.: Nucl. Phys. B
\textbf{1018}(2025)117017; Bhatti, M.Z. et al.: Int. J. Geom.
Methods Mod Phys. \textbf{0}(2025)2550209.

\bibitem{z5} Nasir, M.M.M. et al.: Eur. Phys. J. C \textbf{85}(2025)159.

\bibitem{z6} Javed, F. et al.: Ann. Phys. \textbf{476}(2025)169956.

\bibitem{06h} Rehman, A. et al.:  Eur. Phys. J. C
\textbf{85}(2025)949; Farwa, U. Abass, A. and Yousaf, M.: Nucl.
Phys. B \textbf{1018}(2025)117086.

\bibitem{006h} Yousaf, M. et al.: Chin. J. Phys.
\textbf{97}(2025)1284.

\bibitem{z7} Javed, F. et al.: Nucl. Phys. B \textbf{1018}(2025)117001.

\bibitem{z8} Javed, F. et al.: Ann. Phys. \textbf{482}(2025)170189.

\bibitem{6iii} Yousaf, M. Asad, H. and Rehman, A.: Phys.
Dark Universe \textbf{48}(2025)101888.

\bibitem{z9} Malik, A. et al.: Phys. Dark Universe \textbf{50}(2025)102114.

\bibitem{z10} Rani, S. et al.: Phys. Dark Universe \textbf{47}(2025)101754.

\bibitem{z11} Fatima, N. et al.: Nucl. Phys. B \textbf{1016}(2025)116923.

\bibitem{z12} Rani, S. et al.: Mod. Phys. Lett. A  \textbf{40}(2025)2450213.

\bibitem{7} Katirci, N. and Kavuk, M.: Eur. Phys. J. Plus \textbf{129}(2014)163.

\bibitem{08} Pandya, D.M., Thomas, V.O. and Sharma, R.: Astrophys. Space Sci. \textbf{356}(2015)292.

\bibitem{00024} Roshan, M. and Shojai, F.: Phys. Rev. D \textbf{94}(2016)044002.

\bibitem{00025} Board, C.V.R. and Barrow, J.D.: Phys. Rev. D \textbf{96}(2017)123517.

\bibitem{9} Akarsu, O., Katirci, N. and Kumar, S.: Phy. Rev. D \textbf{97}(2018)024011.

\bibitem{12} Nari, N. and Roshan, M.: Phys. Rev. D
\textbf{98}(2018)024031.

\bibitem{12a} Moraes, P.H.R.S. and Sahoo, P.K.: Phys. Rev. D
\textbf{97}(2018)024007.

\bibitem{12b} Bahamonde, S., Marciu, M. and Rudra, P.: Phys. Rev. D
\textbf{100}(2019)083511.

\bibitem{00028} Ranjit, C., Rudra, P. and Kundu, S.: Ann. Phys. \textbf{428}(2021)168432.

\bibitem{z13} Gul, M.Z. et al.: Phys. Dark Universe \textbf{45}(2024)101537.

\bibitem{z14} Sharif, M. et al.: Phys. Dark Universe \textbf{46}(2024)101606.

\bibitem{z15} Gul, M.Z. et al.: Chin. J. Phys. \textbf{89}(2024)1347.

\bibitem{z16} Sharif, M. et al.: Chin. J. Phys. \textbf{89}(2024)266.

\bibitem{z17} Hashim, I. et al.: High Energy Density Phys. \textbf{57}(2025)101223.

\bibitem{z18} Hashim, I. et al.: Int. J. Geom. Methods Mod. Phys. \textbf{22}(2025)2540049.

\bibitem{029} Chattopadhyay S. et al.: Astrophys. Space Sci. \textbf{353}(2014)27.

\bibitem{030} Jawad, A. and Rani, S.: Astrophys. Space Sci. \textbf{359}(2015)23.

\bibitem{031} Jawad, A. and Chattopadhyay, S.: Astrophys. Space Sci.
\textbf{357}(2015)37.

\bibitem{0034}  Odintsov, S.D., Oikonomou, V.K. and Banerjee, S.: Nucl. Phys. B \textbf{938}(2019)935.

\bibitem{1} Chiba, T., Okabe, T. and Yamaguchi, M.: Phys. Rev. D
\textbf{62}(2000)023511; Kamenshchik, A.Y., Moschella, U. and
Pasquier, V.: Phys. Lett. B \textbf{511}(2001)265.

\bibitem{2} Li, M.: Phys. Lett. B \textbf{603}(2004)5.

\bibitem{2.001} Moradpour, H. et al.:
Eur. Phys. J. C \textbf{78}(2018)6.

\bibitem{2.01} Tavayef, M. et al.: Phys. Lett. B \textbf{781}(2018)195.

\bibitem{2.02} Jahromi, A.S. et al.: Phys. Lett. B \textbf{780}(2018)21.

\bibitem{24} Jawad, A. et al.:  Symmetry \textbf{10}(2018)635.

\bibitem{24.01} Moradpour, H. et al.: Eur. Phys.
J. C \textbf{78}(2018)829.

\bibitem{25} Iqbal, A. and Jawad, A.:
 Phys. Dark Universe \textbf{26}(2019)100349.

\bibitem{26} Maity, S. and Debnath, U.: Eur. Phys. J. Plus
\textbf{134}(2019)514.

\bibitem{26.01} Prasanthi, U.D. and Aditya, Y.: Results Phys. \textbf{17}(2020)103101.

\bibitem{27} Sharma, U.K. and Dubey, V.C.:  Eur. Phys.
J. Plus \textbf{135}(2020)391.

\bibitem{27.01} Bandyopadhyay, T. and Debnath, U.: Mod. Phys. Lett. A \textbf{36}(2021)2150081.

\bibitem{28} Shekh, S.H., Moraes, P.H. and Sahoo, P.K.: Universe \textbf{7}(2021)67.

\bibitem{29} Sardar, A. and Debnath, U.: Mod. Phys. Lett. A \textbf{36}(2021)2150180.

\bibitem{30} Nojiri, S.I., Odintsov, S.D. and Paul, T.: Symmetry \textbf{13}(2021)928.

\bibitem{30.01} Upadhyay, S. and Dubey, V.C.:  Gravit. Cosmol. \textbf{27}(2021)281.

\bibitem{31} Aditya, Y. and Prasanthi, U.D.: Bulg. Astron. J. \textbf{38}(2023)52.

\bibitem{31.01} Aditya, Y., Tejeswararao, D. and Prasanthi, U.D.:
 East Eur. J. Phys. \textbf{1}(2024)85.

\bibitem{z19} Gul, M.Z. et al.: Astron. Comput.
\textbf{52}(2025)100956.

\bibitem{z20} Gul, M.Z. et al.: Eur. Phys. J. Plus \textbf{140}(2025)18.

\bibitem{z21} Sharif, M. et al.: High Energy Density Phys. \textbf{55}(2025)101185.

\bibitem{31.0001} Gaztanaga, E., Bonvin, C. and Hui, L.: J. Cosmol. Astropart. Phys.
\textbf{01}(2017)032.

\bibitem{31.0002} Jiao, K., Borghi, N., Moresco, M. and Zhang, T.J.:
Astrophys. J. Suppl. \textbf{265}(2023)48.

\bibitem{32} Cohen, A.G., Kaplan, D.B. and Nelson, A.E.: Phys. Rev. Lett.
\textbf{82}(1999)4971.

\bibitem{34a} Akrami, Y. et al.: Astron. Astrophys. \textbf{641}(2020)A10.

\bibitem{33} Jawad, A. et al.: Symmetry, \textbf{10}(2018)635.

\bibitem{34} Saleem, R., Ijaz, A. and Waheed, S.: Fortschr. der Phys. \textbf{73}(2025)2300276.

\end{thebibliography}
\end{document}